\documentclass[useAMS,usenatbib,a4paper]{mn2e}

\pdfpagewidth=\paperwidth
\pdfpageheight=\paperheight

\newcommand{\sub}[2]{\ensuremath{#1_{\mathrm{#2}}}}
\newcommand{\super}[2]{\ensuremath{#1^{\mathrm{#2}}}}
\newcommand{\rcore}{\sub{r}{core}}
\newcommand{\unit}[2]{\ensuremath{\textrm{#1}^{#2}}}

\usepackage{graphicx} 
\usepackage{epstopdf} 
\usepackage{amsmath}

\begin{document}

\title[Dark matter interactions in tidal streams]{Enhancements to velocity-dependent dark matter interactions from tidal streams and shells in the Andromeda galaxy} 
\author[R. E. Sanderson, R. Mohayaee, and J. Silk]{Robyn E. Sanderson${}^{1}$, Roya Mohayaee${}^2$, and Joe Silk${}^{2,3}$\\
${}^{1}$Kapteyn Astronomical Institute, P.O. Box 800, 9700 AV Groningen, the Netherlands\\
${}^2$UPMC, CNRS, Institut d'Astrophysique de Paris, 98 bis Bd. Arago, Paris 75014, France\\
${}^3$Physics Department, University
of Oxford, OX1 3RH Oxford, United Kingdom}

\maketitle

\begin{abstract} 
Dark matter substructure around nearby galaxies provides an interesting opportunity for confusion-free indirect detection of dark matter.  We calculate the boost over a
smooth background distribution of dark matter for gamma-ray
emission from dark matter self-annihilations in tidal structure in
M31, assuming a cross-section inversely proportional to the
relative velocities of the dark matter particles as proposed by the Sommerfeld effect.  The low velocity of the material in the structure results in a significant
increase in gamma-ray emission compared to both the background
halo and the predicted emission for a velocity-independent cross
section.  We also calculate the expected signal for {\it Fermi}, for reasonable choices of the dark matter parameters.  We find that for a cross section proportional to $v^{-2}$, the enhancement to the annihilation rate is sufficient to test the velocity dependence of the cross section by spatial correlation with the stellar component of the stream, given sufficient detector sensitivity.
\end{abstract}


\section{Introduction}

Like our own galaxy, the nearby  
Andromeda galaxy (M31) exhibits a wealth of stellar features with an accretion
origin, including dwarf galaxies, tidal streams, and a complex outer
halo structure \citep{2009Natur.461...66M}.  Unlike our  galaxy,
however, we view M31 from the outside, which in some cases facilitates
the study of these substructures.  Many of these tidal features are potential tracers of collisionless dark matter associated with their progenitors, which undergoes similar dynamics to the stars. These features usually have distinctive, asymmetric shapes at large radii from the center of their host.  This paper considers whether dark matter in tidal structures in M31 could provide a possible indirect detection by ultra-sensitive gamma-ray observations in a confusion-free region around this nearby galaxy.

\begin{figure*}
\begin{center} \includegraphics[width=0.9\textwidth]{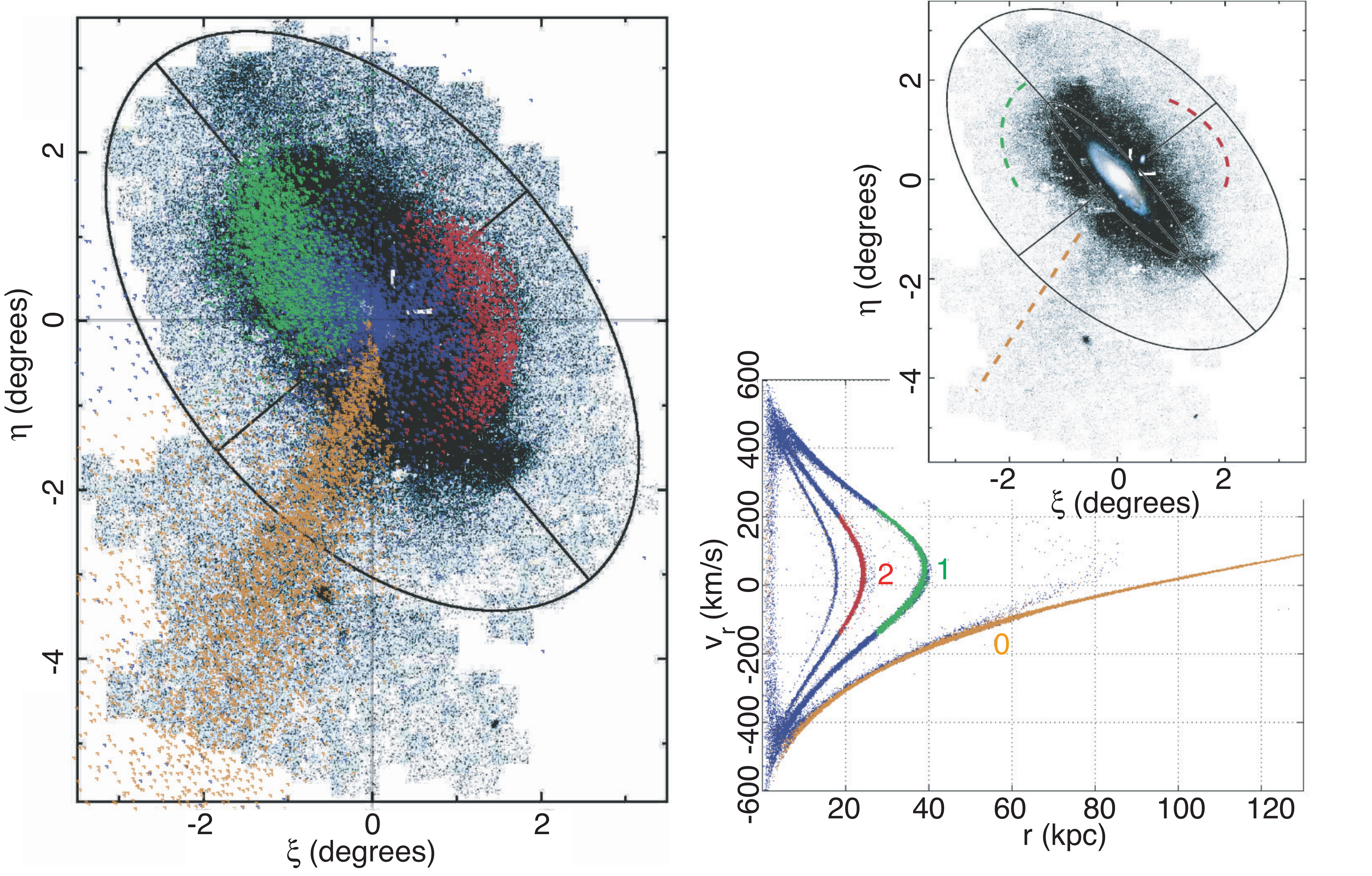}
\caption{Two tidal features noted around the Andromeda galaxy (left
panel, green and red points; top right panel, green and red dashed
lines) correspond to fold catastrophes, or caustics, in individual
phase-wraps of material from a dissolved satellite galaxy on a nearly
radial orbit (right bottom panel).  For this reason we refer to the
shells in this work as Caustic 1 (green) and Caustic 2 (red), in the
order in which they were formed.  Another prominent tidal feature, the
giant stream shown in orange, is the first structure to form in the
merger, and hence is labeled `0'.\label{fig:M31Shells}}
\end{center}
\end{figure*}

One particularly prominent feature around Andromeda is a giant tidal stream that extends nearly radially away from the center of M31, commonly known as the `Giant Stream.'  This feature was first observed by \cite{2001Natur.412...49I} and has since been studied in great detail.  An N-body model of the stream by
\citet{fardal:2006aa} has tentatively connected this stream with two
other tidal features closer to M31's disc, known as the west and
northeast `shelves' because of a relatively abrupt drop in surface
brightness at their edges (Figure \ref{fig:M31Shells}, left panel and
right top panel).  If the three features were all indeed produced in
the same minor merger, the extremely high eccentricity required for
the orbit of the progenitor implies that the `shelves' are in fact
radial fold catastrophes, otherwise known as caustics.  This theory
explains the sharp edges of the shells as the point where in-falling and outgoing streams of material stripped from the progenitor pass each other near the outer radial turning point of their orbits.  Since the motion is nearly
radial, the projection of phase space into the $r-v_r$ plane (Figure
\ref{fig:M31Shells}, right bottom panel) contains nearly all the
information about the dynamics of material in the stream and shells.
A caustic occurs at each point where the phase space stream becomes vertical in this projection, and the various features can thus be placed in chronological order of formation.  This theory also predicts that near the edge of each shelf, the density
will be significantly enhanced, as particles `pile up' near the
radial turning points of their orbits.

\begin{figure}
\begin{center}
\includegraphics[width=0.5\textwidth]{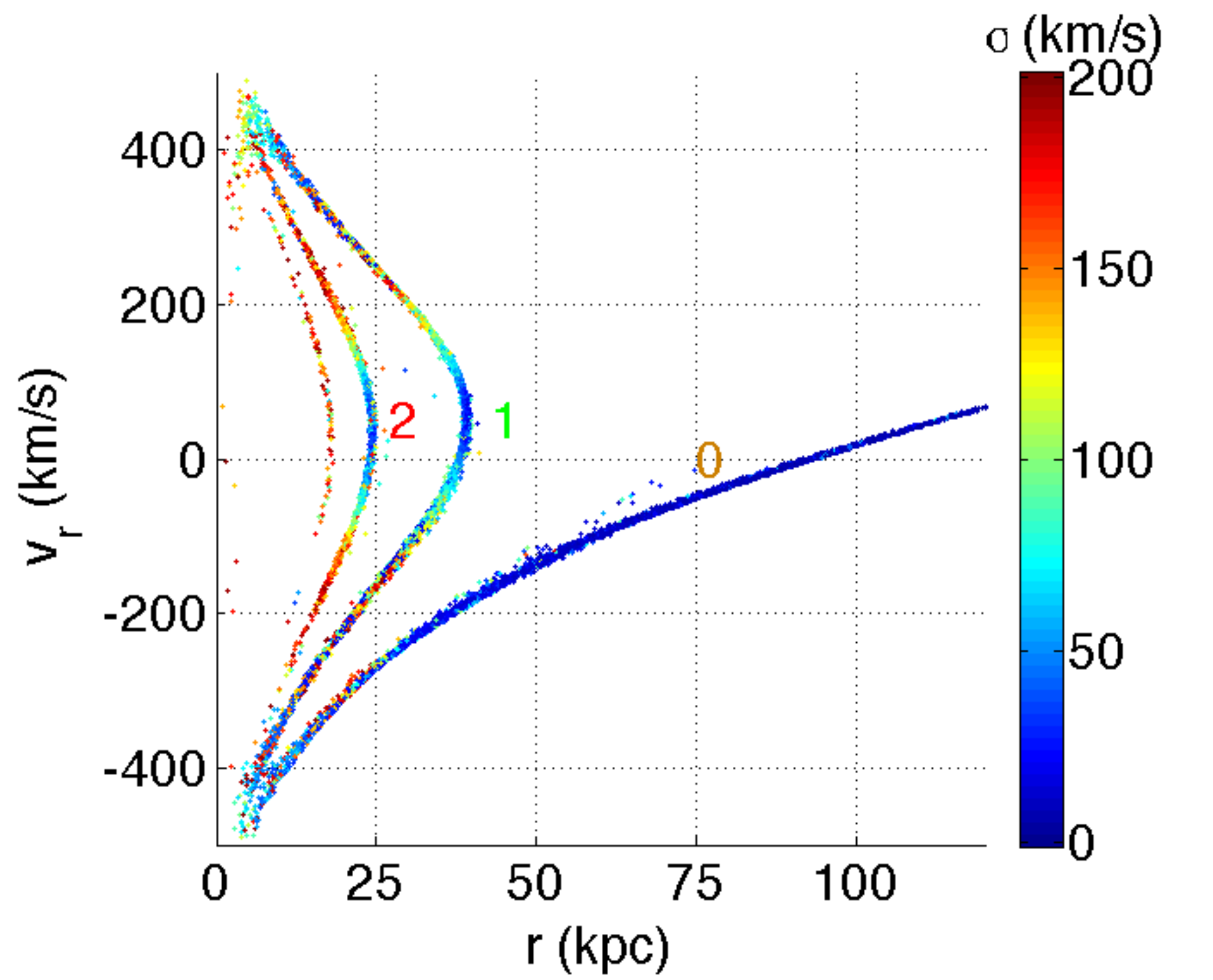}
\caption{Projected phase-space plot of the tidal debris with $\sigma<200$ km/s,
colour-coded by the local velocity dispersion estimated from the N-body model (estimation method described in Section \ref{sec:modelling}).  The cut in velocity dispersion excludes mainly material near the centre of the halo.  The coldest material is found at the edges of the shells and in the tidal tail.  For display purposes, a random selection of one-tenth of the particles are plotted here.}
\label{fig:coldestRegions}
\end{center}
\end{figure}

Inferring the phase space distribution of the material in the shells
and stream from this N-body model also reveals that the relative
velocity of material in the features is extremely low, especially in
the tidal stream and the very edges of the two caustics (Figure
\ref{fig:coldestRegions}).  At the caustic surface and in the stream, the
local relative velocities can be less than 10 km/s (Table \ref{tbl:averageSigmaTidal}).  This is a result of the increasing thinness of the stream in phase space as time passes, an effect sometimes known as `gravitational cooling' \citep{2006MNRAS.366.1217M}. 

\begin{table}
\caption{Mean and minimum estimated velocity dispersions in the features shown in Figure \ref{fig:M31Shells}. \label{tbl:averageSigmaTidal}}
\begin{center}
\begin{tabular}{llll}
&$\langle E(\sigma) \rangle$,  & min($E(\sigma)$), &\\
Feature (colour in Figure \ref{fig:M31Shells}) & km/s & km/s& $N_p$\\
\hline
\hline
Giant Stream 0 (orange) & 24 & 3.6 & 41842\\
Caustic 1 (green) & 70 & 7.3 & 29547\\
Caustic 2 (red) & 84 & 18 & 12263
\end{tabular}
\end{center}
\end{table}%

\begin{figure}
\begin{center} \includegraphics[width=0.5\textwidth]{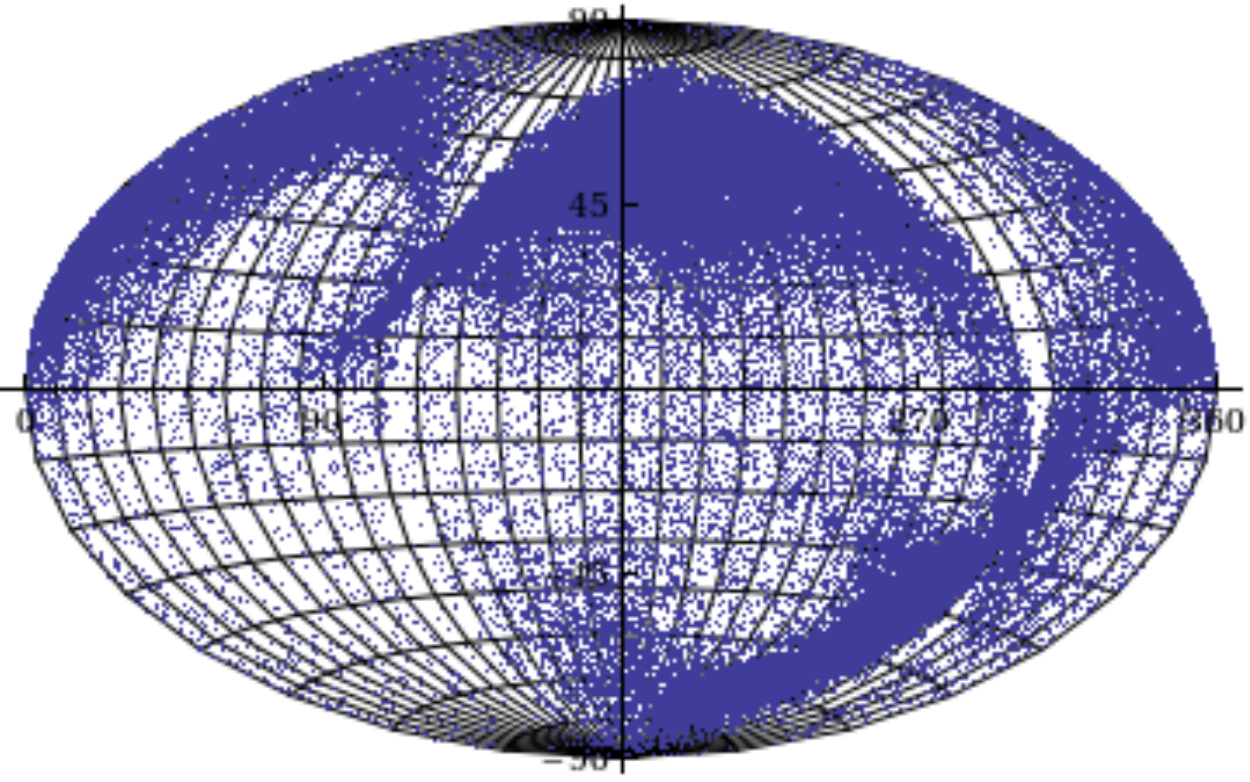}
\caption{The N-body model of the tidal shells in M31, now viewed in
Aitoff projection from a viewpoint at M31's centre.  From this
perspective the shells' sharp edges are virtually
indistinguishable. \label{fig:shellsAitoff}}
\end{center}
\end{figure}

Features of this type, though expected to be fairly common, are as
difficult to detect in the Milky Way as they are straightforward to
find in sufficiently deep images of external galaxies.  The
sharp-edged shells seen in the star-count map of M31 would, when
viewed from within Andromeda, look instead like large amorphous clouds
spread over a huge fraction of the sky (Figure
\ref{fig:shellsAitoff}).  From this vantage point much more
information about the phase space structure of the debris would be
necessary to determine that the shells existed, whereas when viewed
externally in a suitable projection the sharp edges immediately imply a nearly radial orbit for the progenitor.  Thus, the existence of such a structure in M31
represents a unique opportunity to study a system with
well-constrained dynamics, thanks to its distinctive morphology as
viewed from outside, at the closest range possible without full
six-dimensional phase space information for stars in the shells.

The existence of cold, high-density regions at large radii from M31's
centre makes the tidal features an interesting candidate for indirect
detection of dark matter in the Sommerfeld enhancement framework,
where the interaction probability is boosted at low velocities.  The Sommerfeld effect was first introduced to boost the dark matter annihilation signal in order to account for the {\it PAMELA} observations of positrons and {\it HESS} and {\it Fermi} observations of an unpredicted high energy lepton component in the
cosmic rays \citep{2009PhRvD..79a5014A,2010PhRvD..82b3503C,2009PhRvD..79h3523L,2009PhLB..676..133M}. The annihilations of a TeV SUSY WIMP can be boosted by a factor of order 1000, as needed to account for the observed
signal. This interpretation has been criticised on several
grounds. Excessive gamma rays (inverse Compton) and high energy
antiprotons would be produced in the inner galaxy and excessive
radio synchrotron emission in the outer galaxy if the local cold
substructure persists at all galactic radii \citep*{2009ApJ...699L..59B}. The weakness in this critique is that the substructure is likely to be a strong function of
galactic radius. Decrease in substructure concentration factor at
large galactic radius and the effectiveness of tidal disruption at
small galactic radii weaken these constraints significantly. \citet{2011arXiv1107.3546S} have systematically explored this effect, and conclude that our uncertainty about the radial dependence of the substructure contribution means that no strong constraints can be drawn from comparing signals at different Galactic radii.  Any additional information about the expected size of this contribution is therefore quite important for determining the viability of the Sommerfeld model.

The strongest constraint on Sommerfeld boosting has come from
considerations of delayed recombination of the universe following the
last scatterings of the cosmic microwave background radiation at
$z\sim 1000.$ The survival of the model appears marginal for WMAP
constraints, and Planck will soon greatly improve these
limits \citep{2009PhRvD..80b3505G}. However, if the contribution of substructures to the local signal dominates, then these constraints are significantly weakened \citep{2011arXiv1107.3546S}.  Additionally, this line of argument assumes that Sommerfeld boosting, quenched at the local value required to account for {\it PAMELA}, {\it HESS}, and {\it Fermi} observations, applies in a regime where the dark matter is much colder ($\beta\equiv
v/c \sim 10^{-12}$) than in the local halo substructure where
$\beta\sim 10^{-4}.$  This is a huge extrapolation that may not
necessarily be justifiable in terms of general particle physics
models.  We believe it is important to test Sommerfeld boosting in a
regime much closer to the local environment, for example in our galaxy, in ultra-faint dwarf galaxies, and in M31. 

Substructure in M31 can provide precisely such a test if we use old
stars as dark matter tracers. The proposed test involves {\it Fermi}
imaging of very similar substructures (at least in terms of velocity
dispersion) to those invoked locally for the {\it PAMELA} excess.  Without the Sommerfeld enhancement, the material is insufficiently
dense to produce a detectable signal \citep{2010arXiv1006.4165E}, but
with the enhancement signals can be boosted by a factor of up to
$10^{4-5}$ at velocities comparable to those achieved in the tidal
debris in this example.  These boost factors are similar to those expected from local dwarf galaxies \citep{2009PhRvD..79h3525R}.  With this in mind, we calculate in this work
the boost and signal in the {\it Fermi} band from this tidal substructure
assuming that a dark matter component of the unbound substructure
follows the stellar component, as an example of the kind of result one
might expect from tidal debris for this class of dark matter models.
This particular example has then the additional advantage of a
distinctive morphology that could allow it to be easily differentiated
from a smooth dark matter halo. It also occupies an interesting niche between the bound substructures thought to dominate the signal in the Milky Way's outer halo and the more diffuse tidally disrupted substructure that \citeauthor{2011arXiv1107.3546S} propose contributes to the extragalactic gamma-ray background.  In Section \ref{sec:modelling}, we describe the method by which the phase space distribution in M31's halo and tidal substructure were modelled, and
the results of tests for possible bias in our  numerical methods. In
Section \ref{sec:boostfactor} we present results for the boost factor
over the smooth halo as a result of the tidal substructure for
different regimes of Sommerfeld boosting.  In Section
\ref{sec:FermiSignals} we present maps of the flux in the {\it Fermi} band
for two choices of dark matter model and show how those results may be
scaled to other parameter choices.  In Sections \ref{sec:discussion}
and \ref{sec:future} we discuss the results and indicate paths for
future work.

\section{Modelling}
\label{sec:modelling}

The rate $\Gamma$ at which dark matter self-annihilations occur is
proportional to the volume integral of the total squared number
density of dark matter $\sub{n^2}{tot}$, weighted by some function $S(v)$ of the
relative velocity of particles whose form depends on the class of
theories being considered.  In our model, there are two distinct
density distributions that contribute to the total density: the smooth
halo distribution $n_h$ and the tidal structure $n_s$.  The total rate
can thus be separated into three different contributions for ease of
calculation: one from interactions between dark matter particles in
the smooth halo (denoted with a subscript $hh$), one from dark matter
in the tidal structure interacting with dark matter in the smooth halo
(denoted with a subscript $hs$), and one from dark matter particles in
the tidal structure interacting with each other (denoted with a
subscript $ss$):  
\begin{eqnarray} \sub{\Gamma}{tot} &=& \int \sub{n^2}{tot} S(v) dV
\nonumber \\ &=& \int n_h^2 S(\sigma_h) dV + 2 \int n_h n_s S(v_s) dV
+ \int n_s^2 S(\sigma_s) dV \nonumber\\ &\equiv& \sub{\Gamma}{hh} +
\sub{\Gamma}{hs} + \sub{\Gamma}{ss}
\label{eq:boostComponents}
\end{eqnarray} Here we have suppressed the position-dependence of the
arguments for brevity, and denoted the volume element as $dV$.  The
argument to $S(v)$ varies for these three terms.  For
$\sub{\Gamma}{hh}$, the correct relative velocity is the velocity
dispersion $\sigma_h$ of the halo.  For $\sub{\Gamma}{hs}$, the halo
is assumed to have zero mean velocity relative to the debris, so the
mean velocity $v_s$ of shell particles is used.  For
$\sub{\Gamma}{ss}$, the velocity dispersion of particles in the tidal
debris, $\sigma_s$ is used.  The shells and tail have such a low velocity
dispersion that this last term is anticipated to dominate.

\begin{figure}
\begin{center} \includegraphics[width=0.5\textwidth]{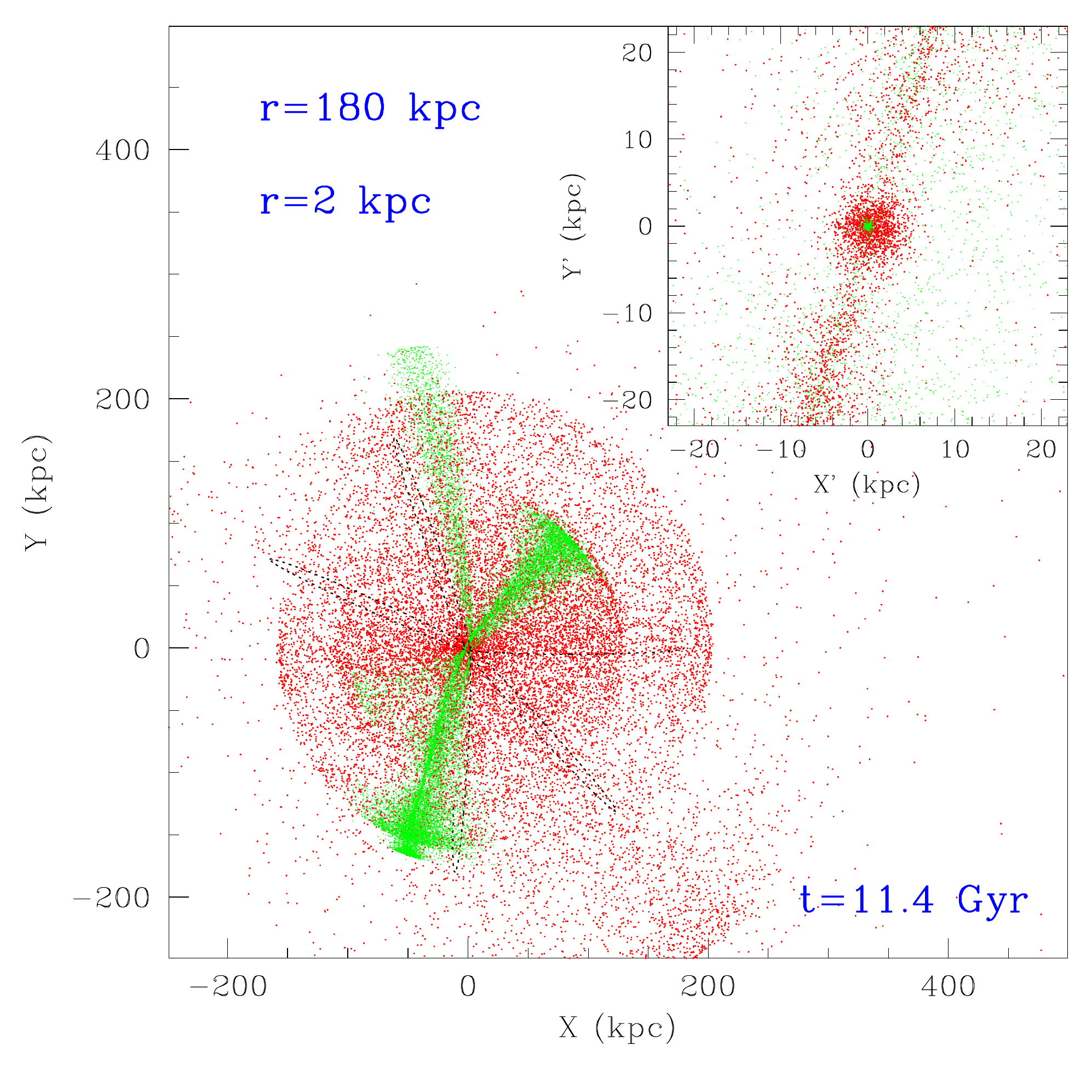}
\caption{A recent simulation of a satellite falling into a Milky Way
type  galaxy \citep[see][for details of the
simulation]{2008ApJ...673..226P}, showing the  creation of shells of both stars (green)
and dark matter (red) in a minor  merger with a nearly radial initial
orbit. The figure shows that star shells  trace very closely the dark
matter shells and despite the difference in initial  distribution, the
dark matter and star shells are at similar radii.  The inset is a
zoom on the satellite which is extremely resilient to the
tides. \label{fig:navarro}}
\end{center}
\end{figure}

To represent the phase space distribution of the material in the tidal
shells, we used the N-body model constructed by
\citet{fardal:2006aa,fardal:2007aa} to match the stellar component of
the tidal debris.  We assumed that the dark component tracks the
stellar component and is of equal mass.  These assumptions are
admittedly an oversimplification but provide a good starting point for
two reasons.  First, the dark matter components of dwarf galaxies are
thought to be more extended and less concentrated than the stellar
component \citep*{2008AN....329..934P} but the starting conditions for
the N-body model locate the satellite deep in the potential of M31, by
which point this extended dark halo would have been tidally stripped
already, leaving only the dark matter within a tidal radius consistent
with the stellar extent of the satellite.  Within this radius, the
dark matter is thought to contribute roughly equally with the stellar
component to the potential of dwarf galaxies, providing some support
for assuming comparable masses for the two components.  This particular N-body model admits a dark matter component up to 2-3 times the total stellar mass (Fardal, private communication). Second, given
the similar initial conditions of the stars and dark matter, the
stellar shells can provide a visible starting point and template for
searches because the radii of the shells, though perhaps not
identical, will be similar.  A justification for this
is seen in simulations that follow both dark and stellar components of
a nearly radial merger, as shown in Figure \ref{fig:navarro}
\citep*[see][for full details of
the simulations]{2008ApJ...672..904P,2008ApJ...673..226P}. The star shells, although formed well after 
the formation of the dark matter shells, clearly trace 
the dark matter shells and form at similar radii.

\subsection{Smooth halo background} 
\begin{figure*} 
\begin{center}
\begin{tabular}{cc}
\includegraphics[width=0.45\textwidth]{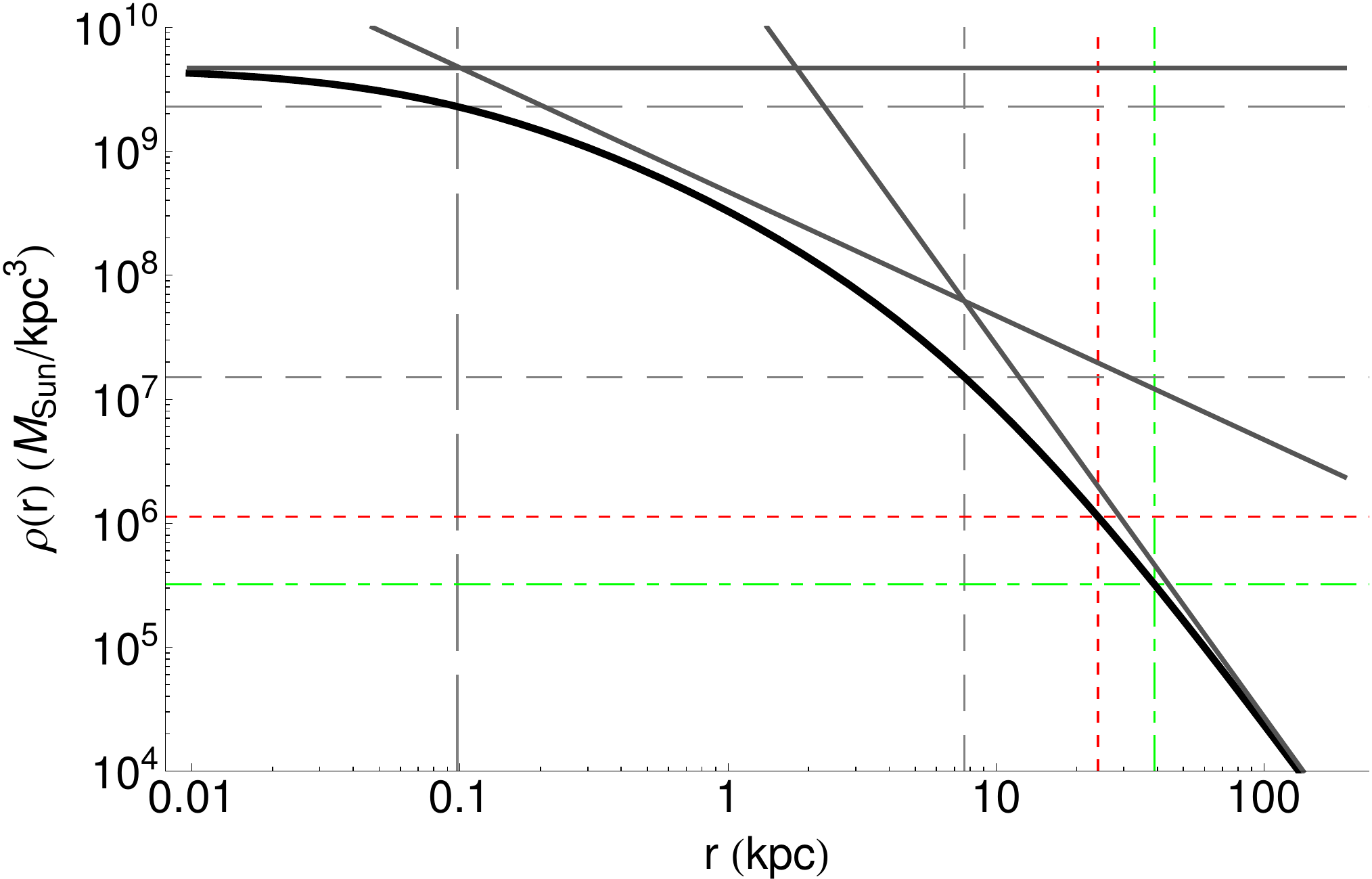} & \includegraphics[width=0.45\textwidth]{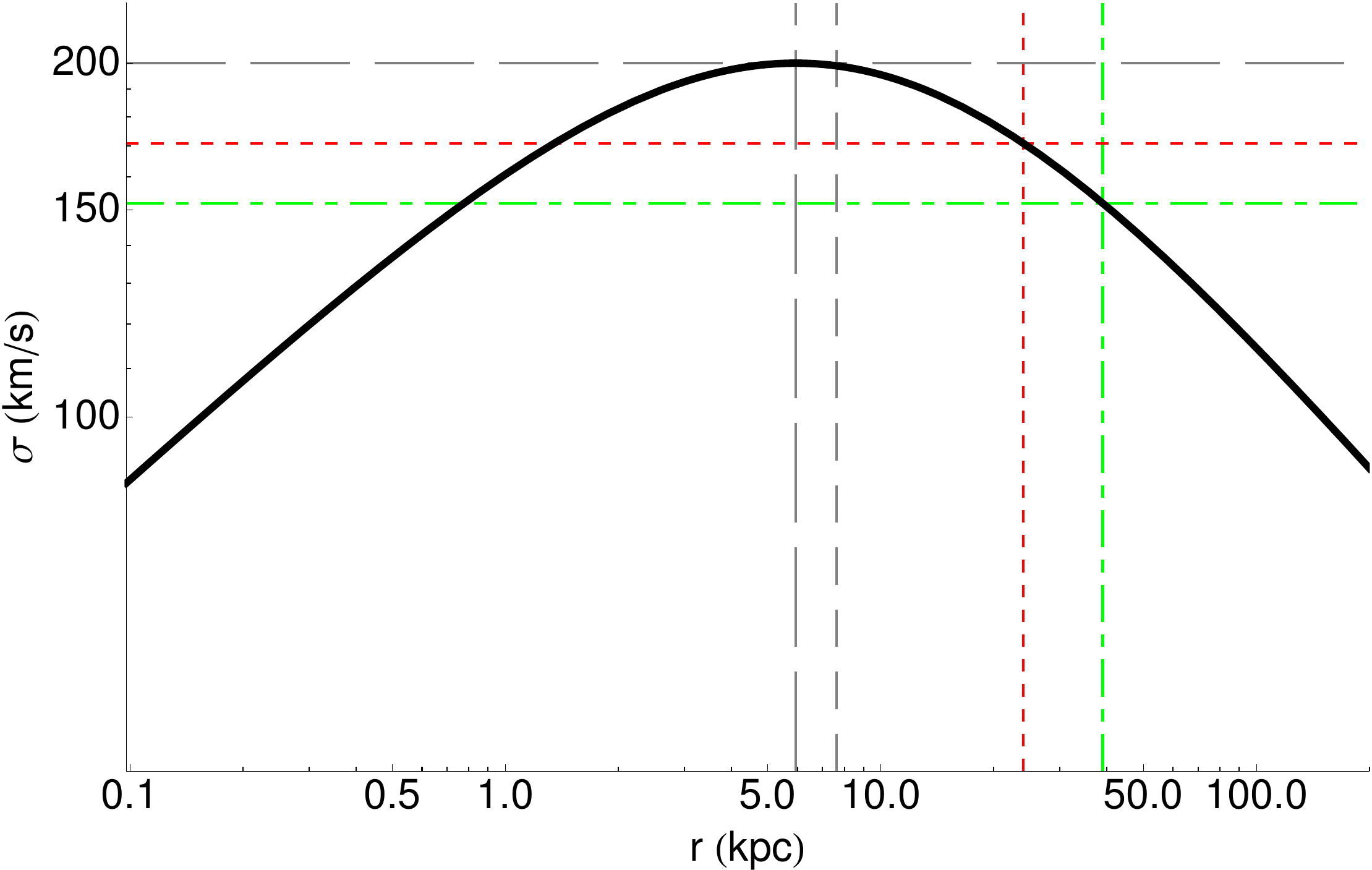}
\end{tabular}
\caption{Left: Density profile of the smooth dark matter halo used in
this work.  The grey solid lines highlight regions where the power law
index is -3 (at large radius), -1 (at intermediate radius), and 0 (at
small radius).  The long-dashed lines indicate $\rcore$, the short-dashed lines
indicate $r_h$, and the dot-dashed (green online) and dotted (red online) solid lines are the approximate
radii of Caustics 1 and 2, respectively. Right: Velocity dispersion profile
of the halo used in this work (solid black line).  The profile peaks
at $\sub{r}{max}$ (long-dashed lines), which is just less than the scale
radius $r_h$ (short-dashed line).  The velocity dispersion of halo material
in the region of the two caustics (noted as in left panel) is more than
10 times larger than the velocity dispersion of material in the
caustics. \label{fig:haloProfiles}}
\end{center}
\end{figure*}

The halo used to model the tidal
structure was also used to calculate both the smooth background signal
and the signal from interaction between dark matter in the halo and
putative dark matter in the tidal shells.  To make the N-body model,
only the density profile $\rho(r)$ was necessary
\citep{geehan:2006aa}; it is represented by the NFW relation
\citep*{navarro:1996aa} with scale radius $r_h$ and scale density
$\rho_{h,0}$, with the addition of a small core radius $\rcore$ to
produce a finite central density:
\begin{equation}
\label{eq:haloDP} \rho_h(r) = \frac{\rho_{h,0}}{[(r+\rcore)/r_h] [1 +
(r+\rcore)/r_h]^2},
\end{equation} as shown in the left panel of Figure
\ref{fig:haloProfiles}.  Note that this halo, used consistently for
both the dark matter annihilation background and the N-body model, has
a concentration of $c=25.5$.  This value is significantly higher than
the concentration of a typical simulated, M31-sized dark halo grown
from cosmological initial conditions, which in most cases is in the
range 8-16 \citep{2010MNRAS.402...21N}.  This high concentration is a
byproduct of the dynamical model and relates to the
uncertainty in the mass of M31's disc (the so-called `disc-halo
degeneracy'); in this work it has the additional effect of producing a
larger than usual background signal from the smooth halo density
distribution in the innermost regions of the halo.  

The velocity dispersion $\sigma(r)$ of the halo was inferred by
analogy with the high-resolution numerical studies of the phase space
structure of cosmological haloes by \citet{2010MNRAS.402...21N}.  These
studies confirmed the pseudo-phase-space-density scaling relation
$\rho/\sigma^3 \propto r^{-15/8}$, proposed by
\citet{bertschinger:1985aa} as a universal relation for cosmological
dark matter haloes, over more than four orders of magnitude in radius.
The density profile of the halo was used to determine the radial
profile of the velocity dispersion in the halo, scaled to a maximum
velocity dispersion determined by rescaling one of the haloes studied
in \citeauthor{2010MNRAS.402...21N}.  The average mean velocity of
particles in the halo is assumed to be zero, and the resulting
velocity dispersion profile is
\begin{equation}
\label{eq:haloSigmaProfile} \sigma_h(r) = \sub{\sigma}{max}
\left(\frac{\rho(r)}{\sub{\rho}{max}}\right)^{1/3}
\left(\frac{r}{\sub{r}{max}}\right)^{5/8}
\end{equation} where it can be shown that
\begin{equation} \sub{r}{max} = \frac{7}{9} r_h \qquad \textrm{and}
\qquad \sub{\rho}{max} = \frac{729}{1792} \rho_h \approx 0.41 \rho_h
\end{equation} The velocity dispersion profile is shown in the right
panel of Figure \ref{fig:haloProfiles}.

Using the analytic expressions for $\rho$ and $\sigma$, we then compute
the quantity $\sub{\Gamma}{hh}$ analytically:
\begin{equation} 
\label{eq:haloRateIntegral}
\sub{\Gamma}{hh} = \frac{1}{m_p^2} \int dV \rho^2_h S(\sigma_h)
\end{equation} 
with $\rho_h$ given by Equation \eqref{eq:haloDP} and
$\sigma_h$ given by Equation \eqref{eq:haloSigmaProfile}.  For
consistency with the $N$-body model, we normalise the expression to
the number density of $N$-body particles by dividing by $m_p^2$.  The integral in Equation \eqref{eq:haloRateIntegral} could be taken over the entire volume of the simulation to estimate the total halo flux, but this value depends very strongly on the choice of a core radius for the halo, which is not constrained by the dynamical model.  This sensitivity, however, is confined to a tiny region right at the centre of the halo, equivalent to the central four pixels or so for {\it Fermi}.  To avoid the strong dependence on a parameter that is so ill-constrained, we instead compare the signal from the halo to that from the tidal structure pixel-by-pixel over the field.  Equation \eqref{eq:haloRateIntegral} is evaluated for each pixel separately over the entire line of sight ($z$) and an area on the sky ($\Delta x,\Delta y$) corresponding to the resolution of {\it Fermi} (about 0.1 degrees, or about 1.4 kpc at the distance of M31).  The boost is calculated by comparing the signal in each pixel from the halo and tidal structure.  The uncertainty about the core radius leads to unreliable estimates of the boost only for the central few pixels, while the region of interest is at larger radii where the mass profile is somewhat better constrained, and where the boost factor does not depend so strongly on the slope of the mass profile.

\subsection{Tidal structure}

The tidal structure, including two shells and the giant stream, is
represented by an N-body realisation based on the model constructed by
\citet{fardal:2007aa}.  Although this model is by no means the single
best fit to the available data, it is at least a local best fit that
provides a plausible dynamical origin for the debris and a valuable
tool for inferring the phase-space structure.  Numerical methods are
necessary to estimate the signal from the tidal structure; here we
describe these methods and their limitations.

The integrated squared density, weighted by a factor of either $1/v$
or $1/v^2$, was estimated using the optimal procedure identified in
\citet{2010arXiv1006.4165E}, with the addition of estimates for the
mean velocity (used to account for interactions between shell and halo
dark matter particles) and the velocity dispersion for material in the
shell.  

Moments of the velocity are calculated as follows.  First the mean
velocity $\mathbf{v}$ at the centre of the current Riemann volume,
located at position $\mathbf{x}$, is estimated using
\begin{equation} \hat{\mathbf{v}}(\mathbf{x}) = \frac{1}{N_s}
\sum_{n=1}^{N_s} \mathbf{v}_n.
\end{equation} Here and in the following, the hat symbol indicates an
estimator that recovers a smoothed field from the discrete N-body
representation.  The average relative speed $v_s(\mathbf{x})$ is then
calculated by taking the magnitude of the mean velocity vector:
\begin{equation} \hat{v}_s(\mathbf{x}) = \sqrt{ \hat{\mathbf{v}}\ \cdot
\hat{\mathbf{v}}}.
\end{equation} The quantity $v_s(\mathbf{x})$ represents the relative
velocity between material in the shell and material in the halo at
point $\mathbf{x}$.  

The mean velocity at the point $\mathbf{x}$ is used to compute the
nine-component, symmetric velocity dispersion tensor $\mathsf{\sigma}^{ij}$,
for the orthogonal directions $\{i,j\} \in \{x,y,z\}$, at the same
position $\mathbf{x}$:
\begin{equation} \hat{\mathsf{\sigma}}^{ij}(\mathbf{x}) = \frac{1}{N_s-1}
\sum_{n=1}^{N_s} \left[v_n^i - \hat{v}^i(\mathbf{x})\right]\left[v_n^j -
\hat{v}^j(\mathbf{x})\right]
\end{equation} 

The average one-dimensional velocity dispersion $\sigma_s^2(\mathbf{x})$
is calculated by summing the three eigenvalues $\hat{\sigma}^2_k$ of
the velocity dispersion tensor estimated with
$\hat{\mathsf{\sigma}}^{ij}(\mathbf{x})$ (the lengths of the orthogonal axes of
the velocity ellipsoid):
\begin{equation} \widehat{\sigma^2}_s = \sum_{k=1}^{3}
\hat{\sigma}^2_k
\end{equation} We determine $\hat{\sigma}_s$ by simply taking the
square root.  This quantity represents the relative velocity between
particles in the shell at point $\mathbf{x}$.

Finally, the two components of $\Gamma$ involving the $N$-body model
are computed using the estimators
\begin{equation} \sub{\hat{\Gamma}}{hs} = \frac{1}{m_p}\sum_{V_i \in
V} V_i \hat{n}_{s,i} \rho_h(\mathbf{x}_i) S(\hat{v}_s)
\end{equation} and
\begin{equation} \sub{\hat{\Gamma}}{ss} = \sum_{V_i \in V} V_i\
\widehat{n^2}_{s,i}\ S(\hat{\sigma}_s),
\end{equation} where the Riemann sum is over the volumes $V_i$ making
up the target volume $V$, and $\hat{n}_i$ and $\widehat{n^2}_i$ are
estimated at the centre $\mathbf{x}_i$ of each Riemann volume as in
\citet{2010arXiv1006.4165E}.  The halo density $\rho_h$ is
evaluated at the centre of each Riemann volume for consistency with the positional accuracy of the density estimates.  Because the Riemann volumes are generally small compared to the
gradient of the halo density profile in the regions of interest, the difference between this method of evaluating $\rho_h$ and the analytic integral over each pixel used to calculate $\sub{\Gamma}{hh}$ should likewise be small.

Recovery of smooth fields from a discrete representation can be
sensitive to various discreteness effects, including the choice of
smoothing number $N_s$ and resolution $N_p$ and the local gradient of the
density, especially the existence of sharp edges in the distribution.  Because of the complexity of the method for determining the velocity dispersion, we used numerical experiments to calculate the bias, variance, and rms error of $\sigma_s$ for several different values of $N_s$, and $N_p$ over a range of velocity dispersions.  We looked for variations with these parameters, as well as those due to edge effects or density gradient (which is high near the caustic).  

For the purposes of the numerical experiments, we define the expectation value of the estimator, $E(\hat{\sigma}_s)$, as the mean of the estimated values of $\sigma_s$ over a given region of the sample: 
\begin{equation}
E(\hat{\sigma}_s) \equiv \frac{1}{\sub{N}{sub}} \sum_{i=1}{\sub{N}{sub}} \hat{\sigma}_s(\mathbf{x}_i),
\end{equation}
where $\sub{N}{sub}$ is the number of particles falling in that region of the realization and $\mathbf{x}_i$ is the position of the $i$th particle, with $\hat{\sigma}_s$ defined as described above.  A ``region" could be the entire sample, in which case $\sub{N}{sub}=N_p$, but we also compared subsets that included and excluded edges or caustics.  We compared $E(\hat{\sigma_s})$ to the input value of the velocity dispersion, $\sub{\sigma}{in}$, by computing the bias $\mathcal{B}$ and variance $\mathcal{V}$ of $E(\hat{\sigma}_s)$, defined as
\begin{equation}
\mathcal{B} \equiv \frac{E(\hat{\sigma}_s)}{\sub{\sigma}{in}} - 1 \qquad \textrm{and} \qquad \mathcal{V} \equiv \frac{1}{\sub{N}{sub} \sub{\sigma^2}{in}} \sum_{i=1}^{\sub{N}{sub}} \left[\hat{\sigma}_s(\mathbf{x}_i) - E(\hat{\sigma}_s)\right]^2
\end{equation}
With these definitions, $\mathcal{B}$ represents the average fractional systematic difference between the input and estimated values of $\sigma_s$, and similarly $\mathcal{V}$ measures the average fractional variation of the estimates from their mean (the ``noisiness" of the estimator).  The square root of $\mathcal{V}$, often referred to as the standard deviation, is a measure of the spread of all the individual estimates of $\sigma_s$.  

One can further quantify the performance of the estimator by combining $\mathcal{B}$ and $\mathcal{V}$ in the root-mean-squared (rmse) error, defined as
\begin{equation}
(\textrm{rmse})^2 \equiv \mathcal{B}^2 + \mathcal{V} = \frac{1}{\sub{N}{sub} \sub{\sigma^2}{in}} \sum_{i=1}^{\sub{N}{sub}} \left[\hat{\sigma}_s(\mathbf{x}_i) - \sub{\sigma}{in}\right]^2
\end{equation}
which includes both error from a noisy estimator (in $\mathcal{V}$) and error from a biased one (in $\mathcal{B}$).  It is important to consider the relative contributions of $\mathcal{B}$ and $\mathcal{V}$ to the rmse, however, so we will discuss all three quantities below.

\begin{figure*}
\begin{center}
\begin{tabular}{cc}
\includegraphics[width=0.45\textwidth]{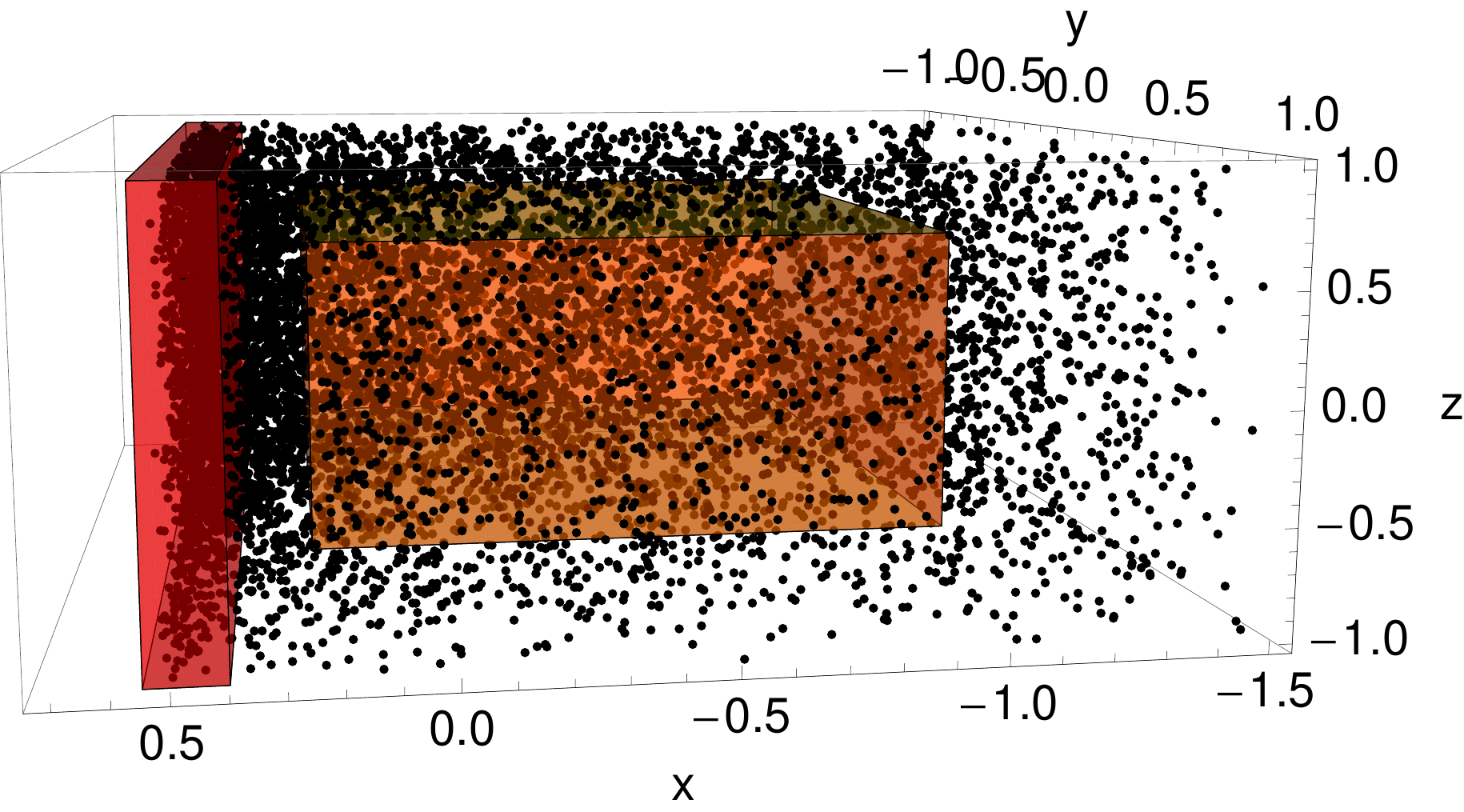} & \includegraphics[width=0.45\textwidth]{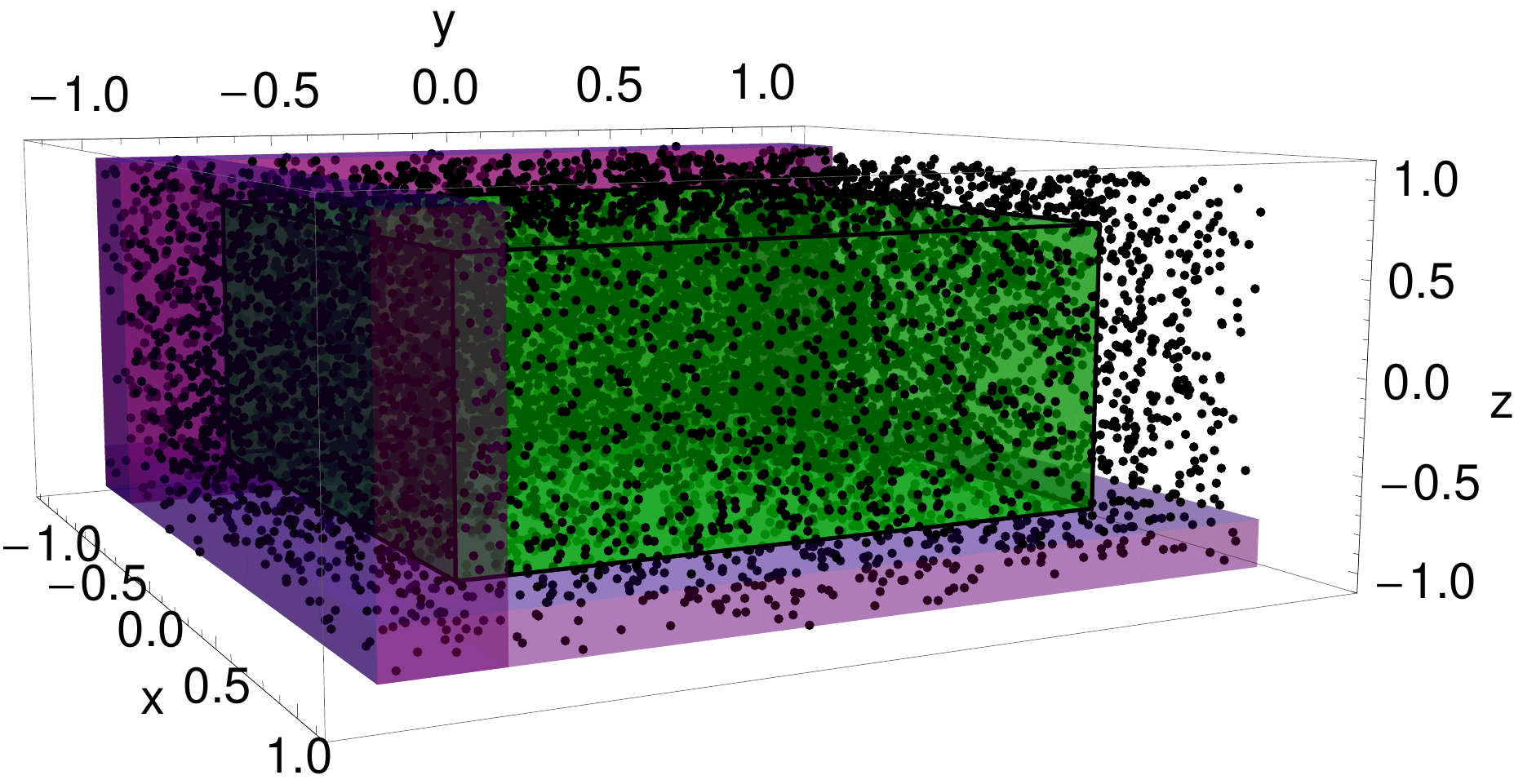}
\end{tabular}
\caption{Regions for testing possible edge and density-gradient
effects in the estimation of the velocity dispersion $\sigma_s$.  The
test compared different regions (edge and center) in two test
distributions: one with a caustic along the $x$ direction (left panel)
and one with uniform density (right panel).  The edge region in the caustic
distribution spanned the caustic surface (red box) while the center
region avoided all the edges and had a small density gradient (orange
box).  The edge region in the uniform distribution covered all the
edges (half are shown in purple) while the central region avoided all
of them (green). \label{fig:EdgeEffectVolumes}}
\end{center}
\end{figure*}

We calculated the bias, variance, and rmse of the expectation values from random realizations of three-dimensional distributions with and without caustics.  The realizations were generated at a range of resolutions between $N_p=10^4$ and $N_p=10^5$, with the number of particles in a given realization chosen from a Poisson distribution with mean $N_p$.  We used a range of velocity dispersions between $10^{-3}$ and $10^{-4}$ and a time unit of 1 to create the caustics.  The caustic width is simply $\delta x = \sigma t$ in our map, so this creates caustics with a thickness of $10^{-3}-10^{-4}$ relative to the units measuring box size as shown in Figure \ref{fig:EdgeEffectVolumes}.  The same velocity dispersion used in the caustic mapping was also used to assign random velocities to each particle: the velocities were reassigned to caustic particles for consistency with the uniform case, and in order to isolate the effect of density gradients on estimates of the velocity dispersion.  The velocity dispersion was calculated at the location of each particle in each sample at $N_s = 10$,  20, and 30.

We tested for edge effects by comparing the estimates of $\sigma_s$
for two types of test distributions with a uniform velocity dispersion: one
with a caustic in it and one with uniform density.  The estimator
described above was used to estimate $\sigma_s$ at the
location of each particle in a given sample.  To test for possible effects
of edges and density gradients, we defined two regions in each sample:
one near the edge of the distribution and one in the centre of the
sample (Figure \ref{fig:EdgeEffectVolumes}).  The edge region in the
caustic distribution is aligned with the caustic surface to probe
possible bias from the high density gradient in the caustic, and spans
several times the scale width of the caustic.  The sizes of the regions are adjusted so they all contain about the same number of particles.  The velocity dispersion was estimated at the locations of particles in a given region using \emph{all} particles in the realization.  If the edges or density gradient affect the estimation of the velocity dispersion, we expect to see a difference in the bias and/or variance of the estimates for regions near the edges or near the caustic compared to regions that exclude the edges.

\begin{figure}
\begin{center} 
\includegraphics[width=0.45\textwidth]{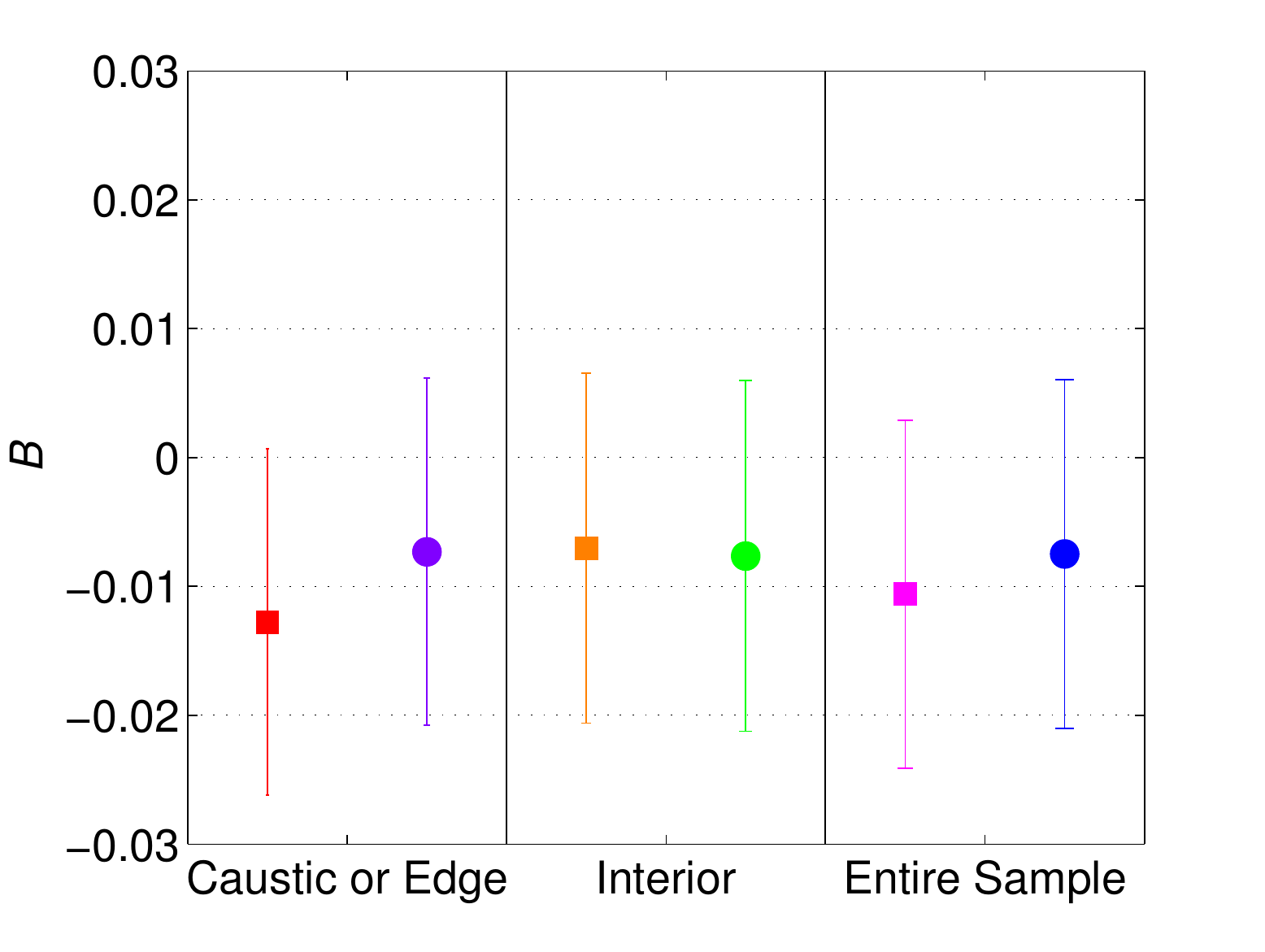}
\caption{The bias (filled points) and rms error (error bars are 1/10 the rmse) of $E(\hat{\sigma}_s)$ do not appreciably differ between the various regions of the sample, or between the samples as a whole.  Print version: squares denote the sample with the caustic; circles denote the uniform-density sample.  Online: The colours in the figure correspond to the regions depicted in Figure \ref{fig:EdgeEffectVolumes}; magenta indicates the entire sample with the caustic and blue indicates the entire uniform-density sample.  \label{fig:BiasVarianceSigma}}
\end{center}
\end{figure}

To illustrate the effect of edges, we present results for $N_s=10$ in a pair of realizations with $N_p=10^5$ (corresponding to $\sub{N}{sub} \sim 10^{4.5}$), since these values of $N_s$, $N_p$, and \sub{N}{sub} are appropriate for the N-body realization of M31.  We found that the rms error on estimates produced by $\hat{\sigma}_s$ was dominated by the large variance resulting from the low value of $N_s$ and was about 14 per cent in all cases.  The bias, which would indicate a systematic error in the estimator, was also independent of the density gradient or the presence of edges in the distribution (Figure \ref{fig:BiasVarianceSigma}).  This rms error corresponds to an uncertainty of about 25 per cent in the Sommerfeld coefficient, which is comparable to our uncertainty about the details of the phase space structure in this tidal debris and certainly less than the uncertainty about the particle physics model of the dark matter.  

The N-body model of the debris uses about 1.3$\times 10^5$ particles
to represent the entire tidal structure, of which about $4\times10^4$
end up in Caustic 1 and $2\times10^4$ end up in Caustic 2.  This level
of resolution is sufficient to resolve the density distribution of the
material if a suitable estimator with a suitably small smoothing
number ($N_s=10$) is used \citep{2010arXiv1006.4165E}, but this does
not guarantee that the velocity structure of the material is
adequately resolved.  To maximise the resolution of the velocity
structure we would like to use the smallest possible smoothing number
to estimate moments of the velocity distribution as well.  

\begin{figure*}
\begin{center}
\begin{tabular}{cc}
\includegraphics[width=0.49\textwidth]{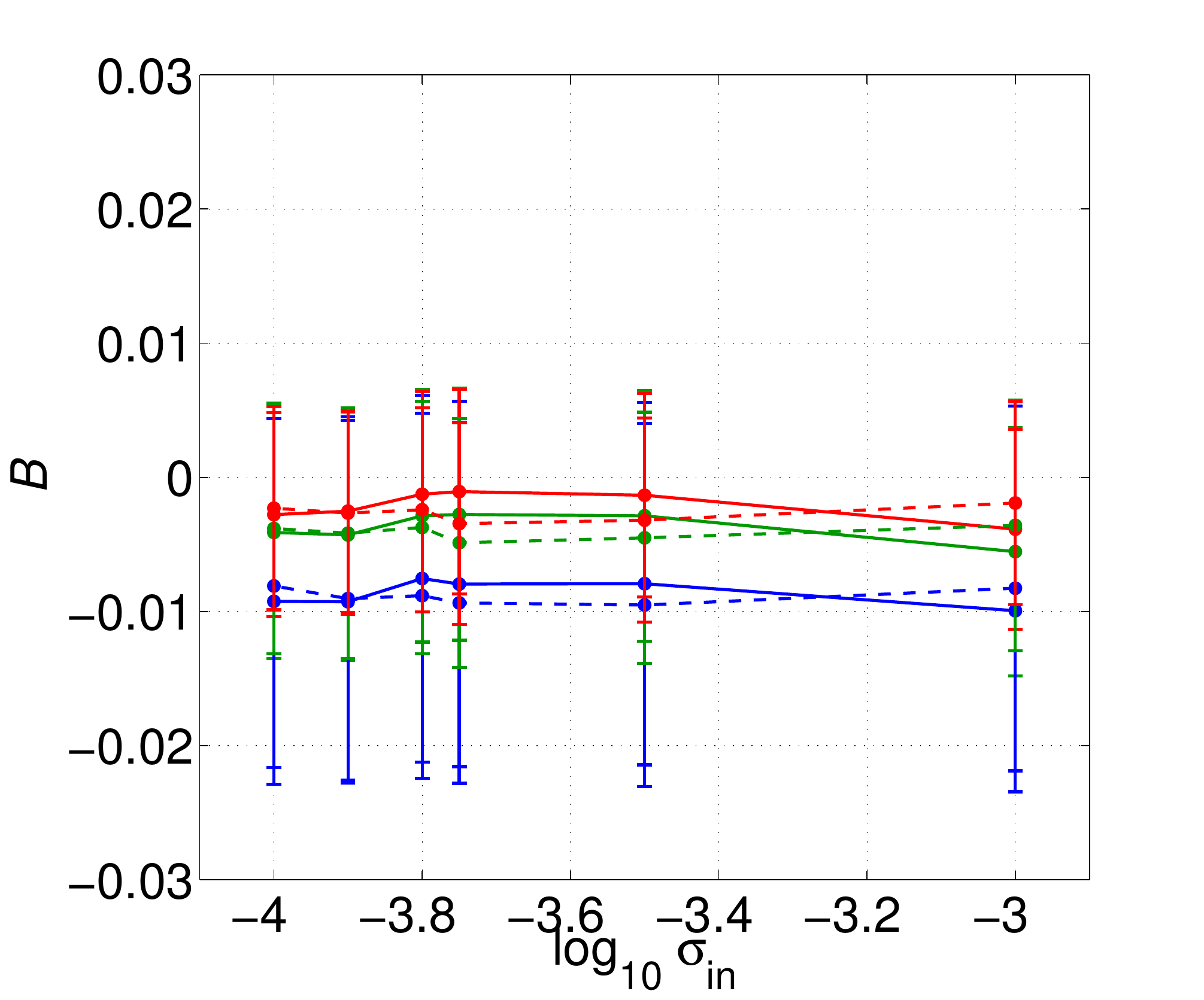} & \includegraphics[width=0.5\textwidth]{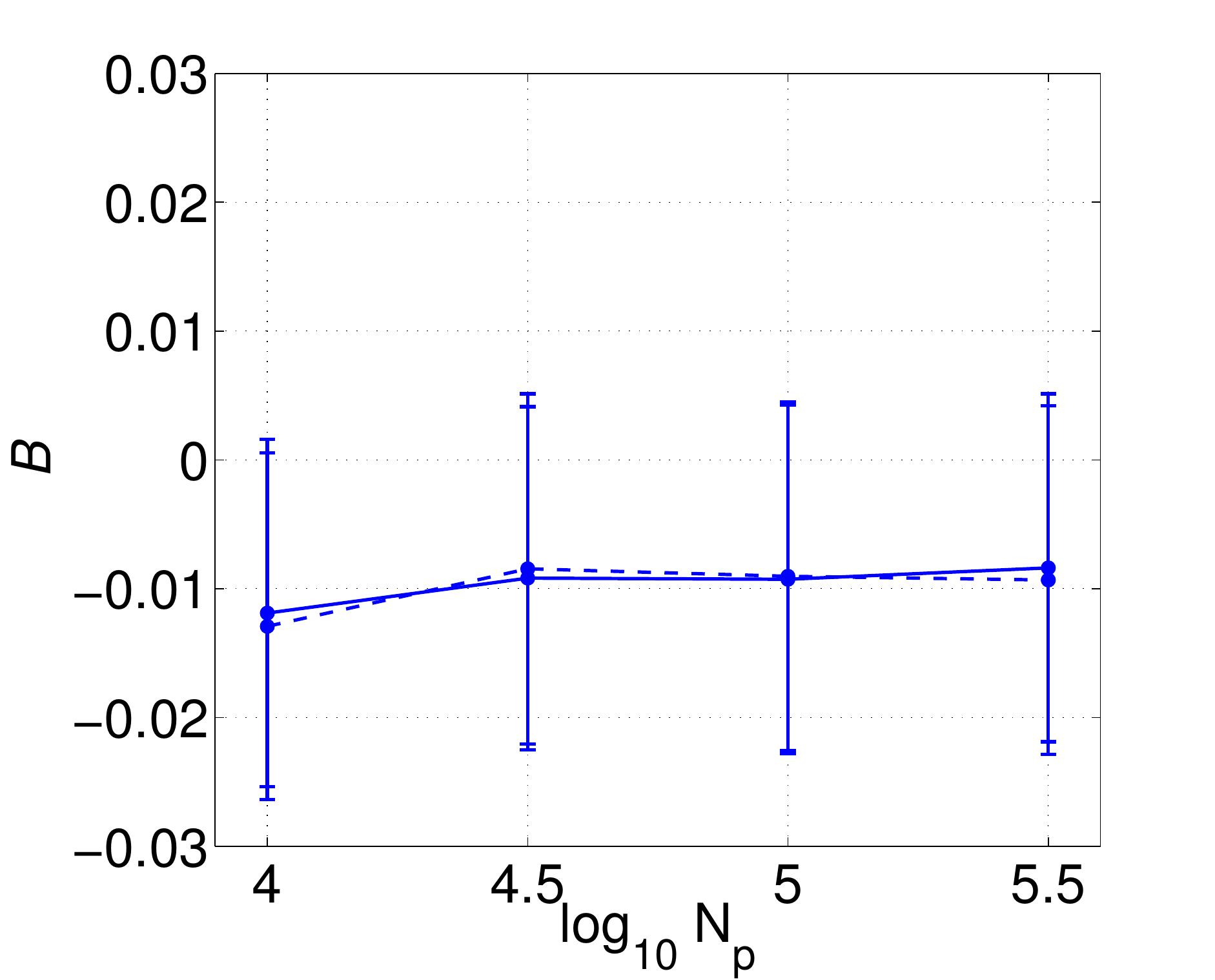}
\end{tabular}
\caption{The ability of the estimator to recover velocity dispersions
is fairly insensitive to most numerical parameters.  The average bias
using $N_s=(10,20,30)$ is shown in (darkest to lightest gray; blue, green, red in online version) in the left panel, for various input values of $\sub{\sigma}{true}$ in kpc
\unit{Myr}{-1}.  Increasing the smoothing number $N_s$ (left panel) or
the resolution $N_p$ (right panel) slightly improves the bias and
rms error regardless of whether the density distribution includes a
caustic (solid lines) or is uniform (dashed lines).  In the right
panel, $N_s=10$ and $N_p$ is varied; above $\log_{10}N_p=4.5$ there is
not an appreciable difference in the bias.  In both panels, the error
bars represent 1/10th of the rms error of the estimator.  \label{fig:velocityEstimatorPerformance}}
\end{center}
\end{figure*}

In order to understand how the choice of smoothing number and the
resolution of the N-body representation affected the sensitivity of
the calculation to small velocity dispersions, we computed the bias
and variance for estimates of $\sigma$ in uniform and caustic density
distributions with different input values of the dispersion
$\sub{\sigma}{true}$, at different smoothing numbers $N_s$, and at
varying resolutions $N_p$.  We find that the estimator can reliably estimate velocity dispersions as small as $10^{-4}$, with no indication that the
bias is dependent on $\sigma$ (Figure \ref{fig:velocityEstimatorPerformance}, left panel).  If the box size is rescaled to the approximate size of the caustics in the N-body simulation of M31, $\sub{\sigma}{in}=10^{-4}$ corresponds to caustics of width $\sim 5$ pc, and velocity dispersions of about 5 km/s.  The caustics in M31 have widths closer to 1 kpc, so their density gradient is always many times smaller than those tested though the minimum velocity dispersions are comparable to this limit (Table \ref{tbl:averageSigmaTidal}).  Choosing a larger $N_s$ does slightly reduce both the bias and the rms error, but will make the estimator less sensitive to small-scale changes in the velocity dispersion.  Increasing the resolution also has only a small effect on the bias and virtually none on the rms error (Figure \ref{fig:velocityEstimatorPerformance}, right panel), which is again dominated by the variance.   Based on these tests we conclude that the current level of resolution of the N-body realisation and the choice of $N_s=10$ will recover adequately
unbiased estimates of the velocity dispersion, sufficient for the required level of accuracy in this work.

\section{Boost factor}
\label{sec:boostfactor}

The boost factor is defined as the enhancement over the smooth halo
provided by the tidal structure:
\begin{equation} B \equiv \frac{\sub{\Gamma}{hs} +
\sub{\Gamma}{ss}}{\sub{\Gamma}{hh}}.
\end{equation}

The boost factor is independent of the normalisation of $S$, the branching ratio, and
other quantities that are determined by the particular particle
physics model of the interaction.  Thus, it usefully isolates the
effect of the different dependences of the cross-section on velocity
without introducing all the complexity of the parameter space of dark
matter models.  We considered two different power laws for $S(v)$ motivated by
previous studies of the Sommerfeld effect \citep[e.g.,][]{2009PhRvD..79a5014A,2009PhRvD..79h3539B,2009PhRvD..79h3523L,2009JCAP...05..016M}: $S(v) \propto 1/v$ and $S(v)\propto 1/v^2$.  For comparison we also consider the
velocity-independent case $S=1$.

\begin{figure*}
\begin{center} \includegraphics[width=\textwidth]{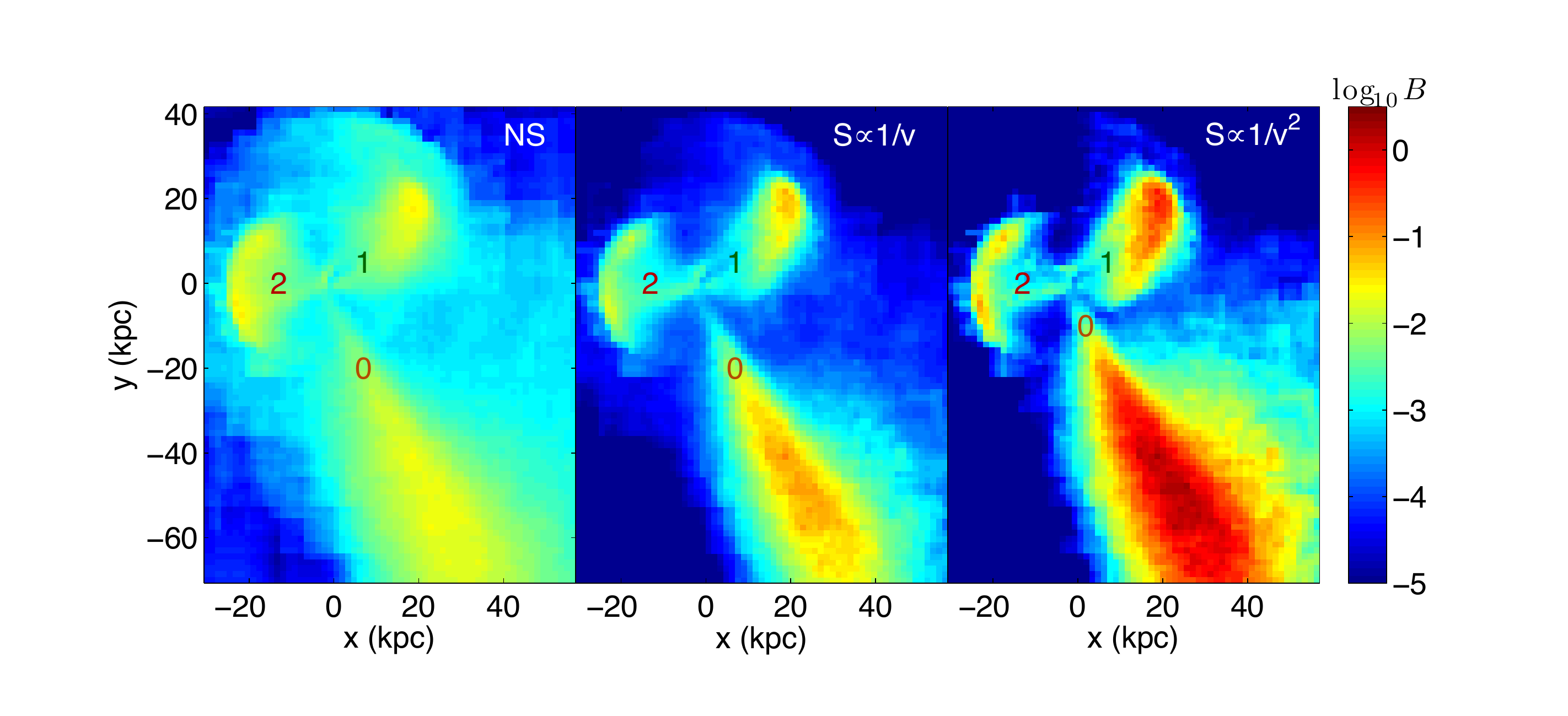}
\caption{Log of the boost factor (ratio of shell emission to smooth
halo emission) for no Sommerfeld enhancement (left panel), $S \propto
1/v$ (center panel), and $S\propto 1/v^2$ (right panel).  The pixel
size is chosen to approximate the resolution capability of the {\it Fermi}
LAT at high energy, as discussed in the text.\label{fig:boosts}}
\end{center}
\end{figure*}

\begin{figure*}
\begin{center} \includegraphics[width=\textwidth]{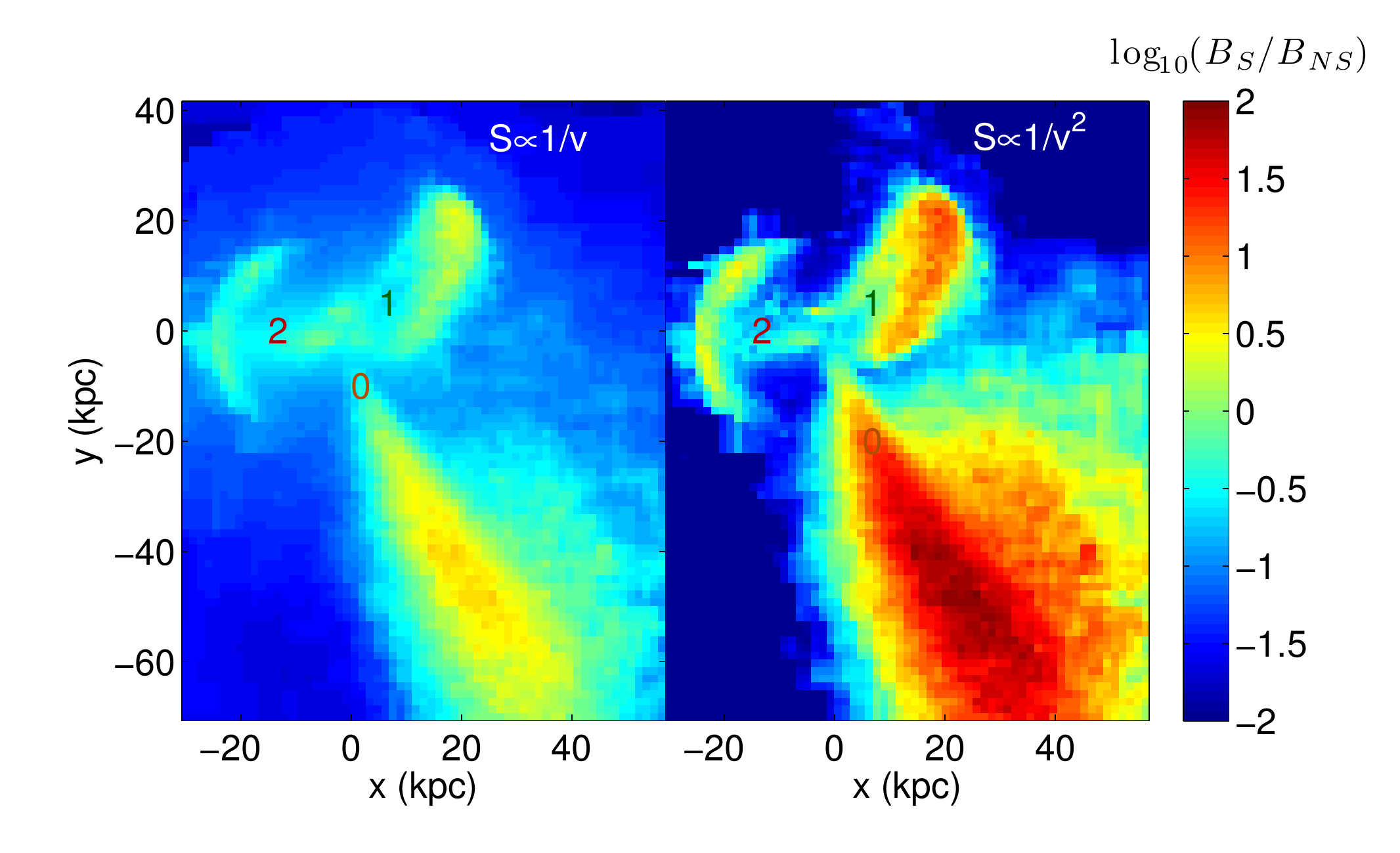}
\caption{Ratio of the boost for velocity-dependent cross sections to
the boost for velocity independent cross sections.  The left panel
shows the enhancement for $S \propto 1/v$ (in other words, the center
panel of Figure \ref{fig:boosts} divided by the left panel) and the
right panel shows the enhancement for $S \propto 1/v^2$ (right panel
of Figure \ref{fig:boosts} divided by left
panel).  \label{fig:boostRatios}}
\end{center}
\end{figure*}

Both velocity-dependent cases (Figure \ref{fig:boosts}, center and
right panels) provide a significant, position-dependent enhancement of
the tidal structure relative to the background and relative to the
velocity-independent case (Figure \ref{fig:boosts}, left panel).  As
expected from Figure \ref{fig:coldestRegions}, the most significant
enhancement is at the edges of the two shells and in the stream, where the density is
highest and the velocity dispersion is lowest.  The enhancement
compared to the velocity-independent case is highly non-linear (Figure
\ref{fig:boostRatios}) because of the correlation between the density
and velocity dispersion, which is a product of the phase-space
streaming and the radial symmetry of the system's dynamics.  

Surprisingly, the tidal stream produces an enhancement that rivals or exceeds
that of the shells.  In retrospect, examination of the right panel of
Figure \ref{fig:M31Shells} and the right panel of Figure
\ref{fig:coldestRegions} shows that the tidal stream is just as cold
as (and perhaps colder than) the shells, its material falling radially
inward in a narrowly collimated and fairly dense band.  The tidal
stream is fairly dense at radii even larger than the edges of the
shells, where the halo is of almost negligible density, leading to an
even larger boost factor.

Of particular interest is the prediction that if the cross-section to
dark matter self-annihilation is proportional to $1/v^2$, the emission
from this tidal structure should be as bright as the background halo
at the edges of the shells and in the densest part of the tidal
stream.  This finding can be used to test the velocity-dependence of
the dark matter cross section if the halo of M31 is detected in
gamma-rays by {\it Fermi}.  If the halo is detected, Figure \ref{fig:boosts}
predicts that zeroth-order departures from a smooth emission
distribution should be observed if the cross-section depends on
$1/v^2$ or higher order, and that these departures should be
correlated with the spatial distribution of tidal structures around
M31.  Likewise, if the cross-section depends on $1/v$, departures
should be observed at the ten percent level, although this may be
beyond the range of current instruments.  These predictions are
independent of the specific model of the dark matter particle, and are
based solely on the assumption of a form for the velocity-dependence
of the cross section.  If no such departures are observed, the class of models with velocity-dependent cross sections of that form
can be ruled out.

\subsection{Spatial correlations}
\label{subsec:correl}
\begin{figure}
\begin{center}
\includegraphics[width=0.5\textwidth]{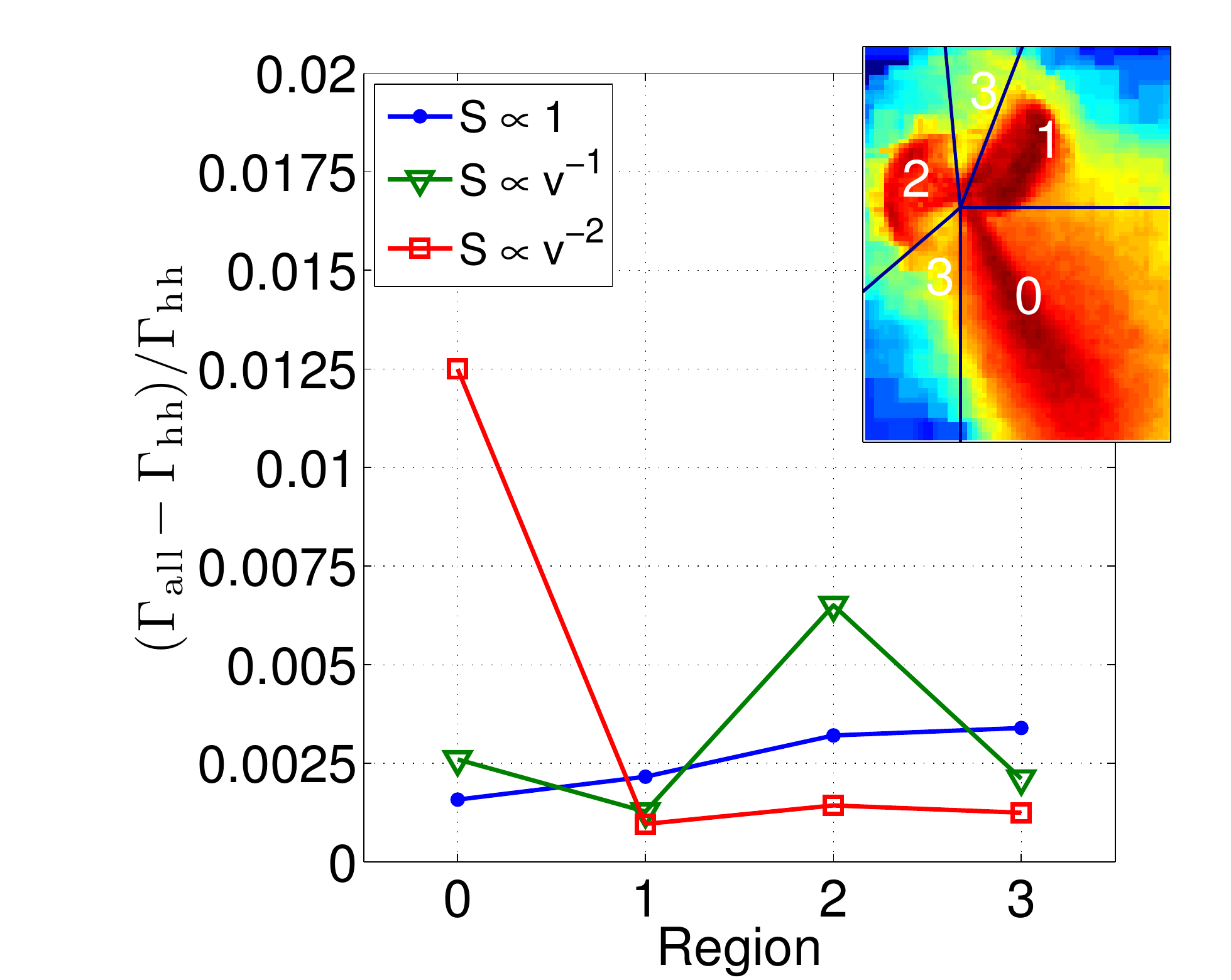}
\caption{Comparing the boost in different regions can reveal deviations from spherical symmetry in the map, even if the signal from the substructure itself is not fully resolved, for some interaction models.  The inset shows the regions over which the sum was calculated. \label{fig:regionalSums}}
\end{center}
\end{figure}

The particular morphology of the tidal features can significantly improve the chances of a detection for a low signal rate, by correlating the stellar map with the gamma-ray map.  Here we demonstrate a coarse version of this by dividing the map into several regions, three centred on a feature and one without significant tidal boosting, and calculating the boost in each of these regions separately (Figure \ref{fig:regionalSums}).  This coarse contrast method can also allow for slight deviations between the dark matter and stellar distributions, though our calculation assumes perfect tracking.  In practice, the boosts would be calculated by fitting a smooth, spherical halo profile to the radially averaged observed distribution of gamma rays, which does not show much deviation for any of the interaction models we considered, and comparing the observed and fitted signal in each pie-shaped region inspired by the arrangement of the tidal debris, which is highly asymmetric.  Including a region assumed to have no boost gives a built-in measurement of the sensitivity of the comparison.

The most prominent feature appears to vary based on the interaction model that is used.  For $S \propto 1/v^2$, the tidal tail is the most prominent, deviating from spherical symmetry at the 1-percent level.  For $S \propto 1/v$, the edge-on shell appears at the 0.5-percent level but the tidal tail is indistinguishable.  This is because the cross-term $\sub{\Gamma}{hs}$ is larger than the shell-shell interaction $\sub{\Gamma}{ss}$ for this case, so that the structure at the smallest radius (i.e. largest halo density) is the brightest.  Without a velocity-dependent cross section, none of the features is distinguishable from the background.

\section{Gamma-ray signal}
\label{sec:FermiSignals}

In order to determine whether the test described in the previous
section could be performed with {\it Fermi}, we estimated the gamma-ray
signal from the halo and tidal substructure for the two forms of $S(v)$ described above.  In this section we discuss the calculation of the signal and its scaling with various parameters.

\subsection{Calculation of the signal}
We follow the notation of \citet*{fornengo:2004aa} as adapted by
\citet{2010arXiv1006.4165E} to calculate the gamma ray signal,
$dN_{\gamma}/dt$, for {\it Fermi}.  As in \citet{fornengo:2004aa}, we
separate the dependence on the phase-space distribution of material
from most of the details of the dark matter particle model:  
\begin{equation}
\label{eq:GammaSignal} \Phi_{\gamma} = \super{\Phi}{C}
\int_{\sub{E}{min}}^{m_\chi} \frac{d\super{\Phi}{P}}{dE_\gamma} A(E_\gamma) {dE_{\gamma}}
dE_{\gamma}
\end{equation} The first term, $\super{\Phi}{C}$, depends only on the
mass and velocity distribution of the dark matter and the
velocity-dependence of the Sommerfeld effect.  This term is
independent of the energy $E_{\gamma}$ of the gamma rays produced in
the interaction:
\begin{equation}
\label{eq:PhiCosmo} \super{\Phi}{C} = \frac{1}{4\pi d^2}
\int_{\mathrm{pixel}} d^3\mathbf{x} \rho^2(\mathbf{x}) S[v(\mathbf{x})]
\end{equation} The signal depends on the distance, $d$, to M31 and on
the local relative velocity $v$ of the dark matter as well as the
local mass density $\rho$.  Of course, the rate at which dark matter
particles interact with each other really depends on the number
density, not the mass density, but since the dark matter mass is
model-dependent there is a corresponding factor of $1/m_{\chi}^2$ in
the second term, $d\super{\Phi}{P}/dE_{\gamma}$, which depends on the
particular model of dark matter being used.  This term describes the
spectrum of gamma rays produced for a given dark matter model:
\begin{equation} \frac{d\super{\Phi}{P}}{dE_{\gamma}} = \frac{\langle
\sigma v \sub{\rangle}{0}}{2 m_{\chi}^2}
\frac{dN_{\gamma}}{dE_{\gamma}}
\end{equation} The cross section $\langle \sigma v \sub{\rangle}{0}$
denotes the value of the cross section without Sommerfeld enhancement.
$dN_{\gamma}/dE_{\gamma}$ is the spectrum of gamma-rays produced in a
particular dark matter model.  

The total signal in a given detector, $dN_{\gamma}/dt$, is calculated
by integrating the spectrum of observed radiation over the energy
range of the detector, weighted by the effective area for the detector
$\sub{A}{eff}$; the only other detector-dependent piece is the lower
limit of the integral $\sub{E}{min}$, the threshold energy for
detection of gamma rays, chosen to be 100 MeV for consistency with the {\it Fermi} sensitivity range.  (The upper limit of the integral is simply
the dark matter mass, as required by energy conservation.) For the
{\it Fermi} LAT, whose effective area is roughly energy-independent above 1
GeV \citep{2009arXiv0907.0626R}, we can calculate the flux
$\Phi_\gamma$ independent of the effective area, since we are
interested mainly in determining whether the structure is above the
detection threshold.  

With this simplification, we find that $\super{\Phi}{P}$ is simply
proportional to the total yield $N_{\gamma}(\sub{E}{min})$ above
$\sub{E}{min}$, so that
\begin{equation}
\label{eq:PhiSUSY} \super{\Phi}{P}(\sub{E}{min}) = \frac{\langle
\sigma v \sub{\rangle}{0}}{2 m_{\chi}^2} N_{\gamma}(\sub{E}{min}).
\end{equation}

\subsubsection{Sommerfeld enhancement}

\citet{2009PhRvD..79a5014A}, \citet{2009PhRvD..79h3523L}, and others
have shown that the Sommerfeld interaction can be easily modelled as a
Yukawa force with coupling constant $\alpha$ and a mediating particle
mass $m_1$.  In this analogy, the solution to the radial Schrodinger
equation with a Yukawa potential exhibits two characteristic behaviours
of the cross section enhancement $S \equiv \langle\sigma v\rangle /
\langle\sigma v\rangle_0$. At very low relative speeds, for resonant
values of the mass ratio $m_{\chi}/m_1$, $S$ is proportional to the
inverse square of the relative speed:
\begin{equation} \super{S}{res} =
\left(\frac{\super{\beta}{*}}{\beta}\right)^2,
\end{equation} where $\beta = v/c$ is the relative velocity of dark
matter particles.  A `low' relative speed is small compared to the
characteristic velocity
\begin{equation} \super{\beta}{*} = \sqrt{\frac{\alpha m_1}{m_\chi}}.
\end{equation} Previous work on the Sommerfeld effect has proposed
values of around $10^{-2}$ for the coupling constant $\alpha$, and a
wide range of $m_{\chi}$ and $m_1$.  We choose $\alpha=1/30$ for this
work, noting that the enhancement only depends on the quantity
$\super{\beta}{*}$ and may thereby be rescaled to any combination of
$\alpha m_1/m_{\chi}$.  The first resonant peak occurs when
$m_1/m_{\chi} \sim 0.2$ \citep{2009PhRvD..79h3523L}.

The N-body realisation gives velocities in the units \unit{kpc}{}
\unit{Myr}{-1} and hence introducing appropriate units and values leads to
the expression
\begin{equation}
\label{eq:SommerfeldFiducial} \super{S}{res} = \super{S}{res}_0
\left(\frac{v}{\unit{kpc}{} \unit{ Myr}{-1}}\right)^{-2}
\left(\frac{\alpha}{10^{-2}}\right)
\left(\frac{m_1/m_{\chi}}{10^{-2}}\right).
\end{equation} where $\super{S}{res}_0 = 9.4$.

Away from resonance, the enhancement takes on the form:
\begin{equation} \super{S}{nr} = \frac{\pi \alpha}{\beta} =
\super{S}{nr}_0 \left(\frac{v}{\unit{kpc}{} \unit{
Myr}{-1}}\right)^{-2} \left(\frac{\alpha}{10^{-2}}\right),
\end{equation} where $\super{S}{nr}_0=9.6$.  This form is also valid
at intermediate values of $\beta$ between $\super{\beta}{*}$ and
$\alpha$.

\subsubsection{Phase-space distribution factor} With an N-body
realisation, the integral in Equation \eqref{eq:PhiCosmo} is
calculated in terms of the number density $n$ for N-body particles of
mass $m_p$, so that:
\begin{equation} \super{\Phi}{C} = \frac{\super{S}{res}_0 m_p^2}{4 \pi
d^2} \left(\frac{\alpha}{10^{-2}}\right)
\left(\frac{m_1/m_{\chi}}{10^{-2}}\right) \int d^3\mathbf{x}\
n^2(\mathbf{x}) \left[\frac{v(\mathbf{x})}{\unit{kpc}{} \unit{
Myr}{-1}}\right]^{-2}
\end{equation} for the resonant interaction and similarly for the
non-resonant regime.  The number density $n$ is likewise in units of
$\unit{kpc}{-3}$ and the volume element $d^3\mathbf{x}$ in units of
$\unit{kpc}{3}$ and hence the result of the numerical integration returned
by the density estimator is
\begin{equation} \frac{\Gamma^{v^2}}{\unit{kpc}{-5} \unit{ Myr}{2}}
\equiv \int \left(\frac{d^3\mathbf{x}}{\unit{kpc}{3}}\right)
\left[\frac{n^2(\mathbf{x})}{\unit{kpc}{-6}}\right]
\left[\frac{v(\mathbf{x})}{\unit{kpc}{} \unit{ Myr}{-1}}\right]^{-2}
\end{equation} on resonance and
\begin{equation} \frac{\Gamma^{v}}{\unit{kpc}{-4} \unit{ Myr}{}}
\equiv \int \left(\frac{d^3\mathbf{x}}{\unit{kpc}{3}}\right)
\left[\frac{n^2(\mathbf{x})}{\unit{kpc}{-6}}\right]
\left[\frac{v(\mathbf{x})}{\unit{kpc}{} \unit{ Myr}{-1}}\right]^{-1}
\end{equation} off resonance.

Using the distance $d=785$ kpc to M31 and the particle mass $m_p
\sim10^4\ \textrm{M}_{\odot}$ from the N-body simulation, and converting to
standard units for the quantity $\super{\Phi}{C}$, we obtain the
phase-space-dependent factor:
\begin{eqnarray}
\label{eq:PhiCosmoFiducialRes} \super{\Phi}{C, res} &=&
\super{\Phi}{C, res}_0 \left(\frac{d}{785 \unit{ kpc}{}}\right)^{-2}
\left(\frac{\alpha}{10^{-2}}\right)
\left(\frac{m_1/m_{\chi}}{10^{-2}}\right) \nonumber \\ &&\times \left(
\frac{m_p}{10^4\ \textrm{M}_{\odot}} \right)^2
\left(\frac{\Gamma^{v^2}}{\unit{kpc}{-5} \unit{ Myr}{2}} \right), 
\end{eqnarray} where $\super{\Phi}{C, res}_0 = 1.75 \times 10^{-13}
\unit{ GeV}{2} \unit{ kpc}{} \unit{ cm}{-6}$, for the resonant
interaction and
\begin{eqnarray}
\label{eq:PhiCosmoFiducialNonRes} \super{\Phi}{C, nr} &=&
\super{\Phi}{C, nr}_0 \left(\frac{d}{785 \unit{ kpc}{}}\right)^{-2}
\left(\frac{\alpha}{10^{-2}}\right)  \nonumber \\ &&\times \left(
\frac{m_p}{10^4\ \textrm{M}_{\odot}} \right)^2
\left(\frac{\Gamma^{v}}{\unit{kpc}{-4} \unit{ Myr}{}} \right), 
\end{eqnarray} where $\super{\Phi}{C, nr}_0 = 1.79 \times 10^{-13}
\unit{ GeV}{2} \unit{ kpc}{} \unit{ cm}{-6}$, for the non-resonant
interaction.

\subsubsection{Particle physics factor} The particle-physics factor in
the flux can be written as
\begin{equation}
\label{eq:PhiSUSYFiducial} \super{\Phi}{P}(\sub{E}{min})   =
\super{\Phi}{P}_0 N_{\gamma}(\sub{E}{min}) \left(\frac{\langle \sigma
v \sub{\rangle}{0}}{3 \times 10^{-26} \unit{ cm}{3} \unit{
s}{-1}}\right) \left(\frac{m_{\chi}}{10\unit{ GeV}{}}\right)^{-2} 
\end{equation} where $\super{\Phi}{P}_0 = 4.63 \times 10^{-7} \unit{
cm}{4}\unit{ kpc}{-1}\unit{ GeV}{-2}\unit{ s}{-1}$.  The gamma ray
yield $N_{\gamma}$ is usually of order 1 per annihilation or less,
depending on the dark matter model \citep{2009JCAP...05..016M}.

\subsubsection{Complete expression} Combining Equation
\eqref{eq:PhiCosmoFiducialRes} or \eqref{eq:PhiCosmoFiducialNonRes}
with and \eqref{eq:PhiSUSYFiducial} gives the master equation for
$\Phi_{\gamma}$ including all the scalings:
\begin{eqnarray}
\label{eq:FermiSignalRes} \super{\Phi_{\gamma}}{res} &=& 8.1 \times 10^{-20}
N_{\gamma} \unit{ cm}{-2} \unit{ s}{-1}\ \left(\frac{m_{\chi}}{10
\unit{ GeV}{}}\right)^{-2}  \nonumber  \\ &&\times
\left(\frac{m_1/m_{\chi}}{10^{-2}}\right)
\left(\frac{\alpha}{10^{-2}}\right) \left(\frac{\langle \sigma v
\rangle_0}{3\times 10^{-26} \unit{ cm}{3}\unit{ s}{-1}}\right)
\nonumber  \\ &&\times  \left(\frac{m_p}{10^4\ \textrm{M}_{\odot}} \right)^2
\left(\frac{d}{785 \unit{ kpc}{}}\right)^{-2}
\left(\frac{E(\hat{\Gamma}^{v^2})}{\unit{kpc}{-5} \unit{
Myr}{2}}\right)
\end{eqnarray} for the resonant process, and
\begin{eqnarray}
\label{eq:FermiSignalNonRes} \super{\Phi_{\gamma}}{nr} &=& 8.3 \times 10^{-20}
N_{\gamma} \unit{ cm}{-2} \unit{ s}{-1}\ \left(\frac{m_{\chi}}{10
\unit{ GeV}{}}\right)^{-2}  \nonumber  \\ &&\times
\left(\frac{\alpha}{10^{-2}}\right) \left(\frac{\langle \sigma v
\rangle_0}{3\times 10^{-26} \unit{ cm}{3}\unit{ s}{-1}}\right)
\nonumber  \\ &&\times  \left(\frac{m_p}{10^4\ \textrm{M}_{\odot}} \right)^2
\left(\frac{d}{785 \unit{ kpc}{}}\right)^{-2}
\left(\frac{E(\hat{\Gamma}^{v})}{\unit{kpc}{-4} \unit{ Myr}{}}\right)
\end{eqnarray} for the non-resonant process. The comparable expression
without the Sommerfeld boost is \citep{2010arXiv1006.4165E}
\begin{eqnarray}
\label{eq:FermiSignalNS} \super{\Phi}{NS}_{\gamma} &=&
9.6\times10^{-22} N_{\gamma} \unit{ cm}{-2}\unit{ s}{-1} \nonumber \\
&& \times  \left(\frac{m_{\chi}}{10 \unit{ GeV}{}}\right)^{-2}
\left(\frac{\langle \sigma v \rangle_0}{3\times 10^{-26} \unit{
cm}{3}\unit{ s}{-1}}\right) \nonumber \\ &&\times
\left(\frac{m_p}{10^4\ \textrm{M}_{\odot}} \right)^2 \left(\frac{d}{785 \unit{
kpc}{}}\right)^{-2}
\left(\frac{E(\sub{\hat{\Gamma}}{NS})}{\unit{kpc}{-3}}\right)
\end{eqnarray}

\subsection{Results for the {\it Fermi} band} 
\begin{figure*}
\begin{center} \includegraphics[width=\textwidth]{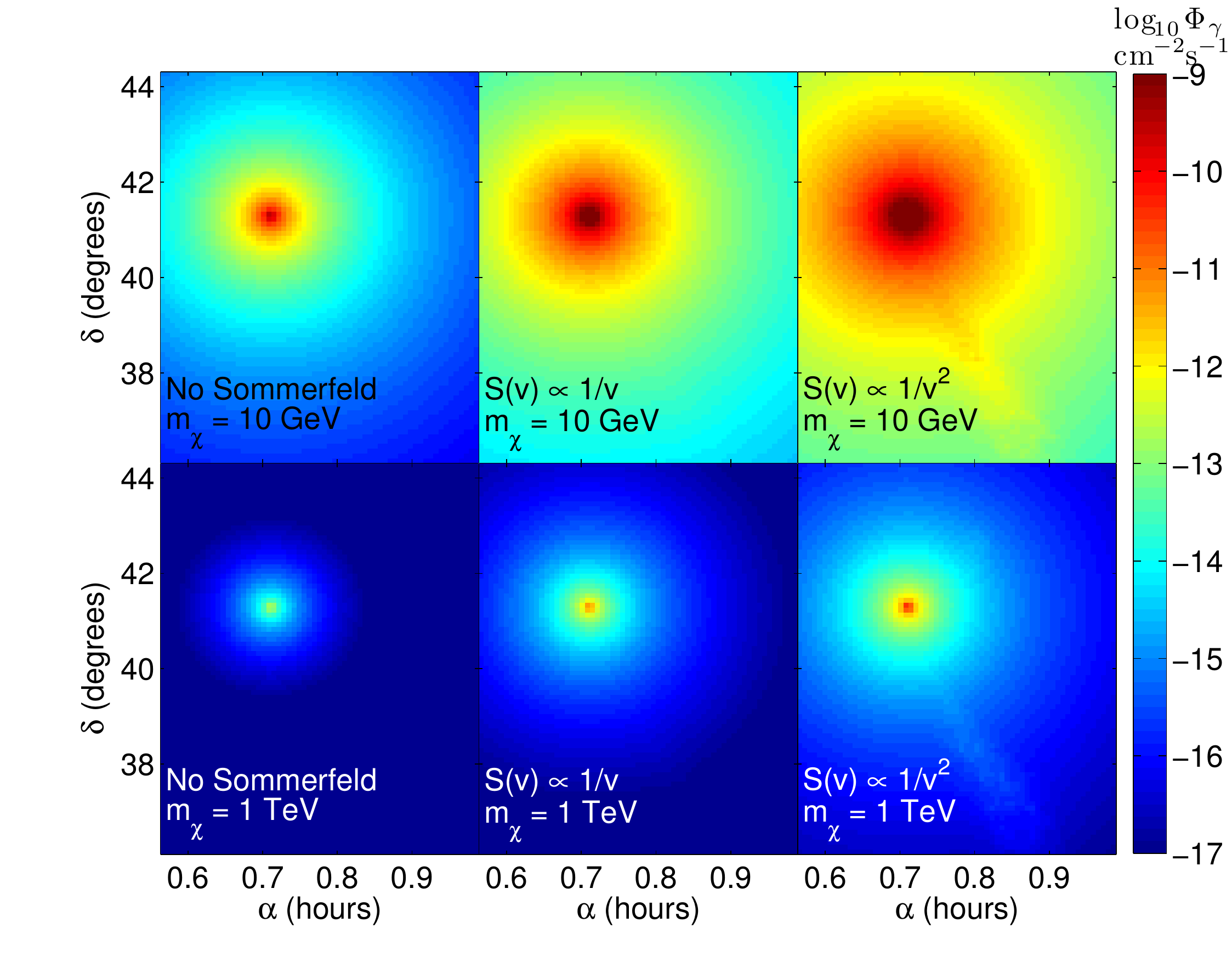}
\caption{Logarithmic map of predicted gamma-ray flux in the {\it Fermi} band
($E_{\gamma}>100$ MeV) for various models of the dark matter
interaction.  The label `no Sommerfeld' indicates direct annihilation
to standard-model particles without an intermediate step.  We consider
the low-mass case $m_{\chi}=10$ GeV with a Sommerfeld-like boost for
comparison, even though the standard mechanism for the Sommerfeld
enhancement is invalid for such a low dark matter mass.  We also take
a typical case for Sommerfeld enhancement with $m_{\chi}=1$
TeV.\label{fig:FermiMap}}
\end{center}
\end{figure*}

Using Equations
\eqref{eq:FermiSignalRes}, \eqref{eq:FermiSignalNonRes} and
\eqref{eq:FermiSignalNS}, we produce maps of the total flux $\Phi_\gamma$ in
gamma-rays in the {\it Fermi} band, including both the halo and the
substructure, for several scenarios.  Figure \ref{fig:FermiMap}
compares the results for resonant, non-resonant, and non-Sommerfeld
cases for two different choices of $m_{\chi}$: 10 GeV and 1 TeV.  The
former is optimistic but realistic for models with no Sommerfeld
boost; the latter is characteristic for models with a Sommerfeld
boost.  We include results at 10 GeV for Sommerfeld-like boosts for
completeness, although a particle model for such an enhancement at low
$m_{\chi}$ does not exist to our knowledge.  However, we do note that \cite{2011arXiv1107.3546S} point out that mediating particles with masses even lower than a few GeV (the lowest considered here) cannot be ruled out by current measurements given the uncertainty about the distribution of substructure in the Galaxy, so this panel may yet be relevant. 

The case $S \propto 1/v$ is brighter than $S = 1$, but only the halo
is visible in both cases: the tidal features are below the smooth
emission by several orders of magnitude.  In Table
\ref{tbl:fermiSignals} we see that the cross-interaction signal
$\sub{\Phi}{hs}$ is larger than the signal $\sub{\Phi}{ss}$ from
interactions within the debris, so that overall the signal from the
substructure scales only linearly with the substructure density at
leading order.  However, the structure may still be marginally detectable using the coarser test described in Section \ref{subsec:correl} given sufficient sensitivity to detect the halo at the appropriate radii. 

In the case where $S \propto 1/v^2$, the
enhancement is non-linear enough in both $\rho$ and $\sigma$ that
although the center of the halo is still the brightest part of the
structure, the tidal features stand out above the halo at their radii.
Table \ref{tbl:fermiSignals} shows that in this case
$\sub{\Phi}{ss}>\sub{\Phi}{hs}$, indicating that for this case the
leading-order signal really scales with the square of the substructure
density.  This structure is still at least an order of magnitude below
{\it Fermi}'s current sensitivity regime, but if such a sensitivity were
achieved, a search for deviations from spherical symmetry in the
gamma-ray emission would be able to test the velocity-dependence of
the interaction cross-section.

We also note that the values of $\sub{\Phi}{hh}$ in the table depend primarily on the inner slope of the mass profile as discussed in Section \ref{sec:modelling}, so the fact that they exceed the measured signal from M31 is merely a reflection of the uncertainty of this parameter.

\begin{table}
\begin{center}
\caption{Total flux from the halo and tidal structure, compared to the
recent {\it Fermi} detection of M31. \label{tbl:fermiSignals}}
\begin{tabular}{p{0.5in}ll} & $\Phi_{\gamma}$, $m_{\chi}=10$ GeV, &
$\Phi_{\gamma}$, $m_{\chi}=1$ TeV,\\ Source & \unit{cm}{-2}\unit{
s}{-1} & \unit{cm}{-2}\unit{ s}{-1}\\ \hline \hline
\super{\sub{\Phi}{hh}}{NS} & $1.6 \times 10^{-9}$ & $1.1 \times
10^{-12}$ \\ \super{\sub{\Phi}{hs}}{NS} & $3.8 \times 10^{-12}$ & $2.7
\times 10^{-15}$ \\ \super{\sub{\Phi}{ss}}{NS} & $4.8 \times 10^{-14}$
& $3.4 \times 10^{-17}$ \\ \hline $\super{\sub{\Phi}{hh}}{nr}$ & $5.0 \times
10^{-8}$ & $2.4 \times 10^{-11}$ \\ $\super{\sub{\Phi}{hs}}{nr}$ & $8.2 \times
10^{-11}$ & $4.0 \times 10^{-14}$ \\ $\super{\sub{\Phi}{ss}}{nr}$ & $5.8 \times
10^{-12}$ & $2.8 \times 10^{-15}$ \\ \hline $\super{\sub{\Phi}{hh}}{res}$ &
$2.9 \times 10^{-7}$ & $1.4 \times 10^{-10}$ \\ $\super{\sub{\Phi}{hs}}{res}$
& $1.5 \times 10^{-10}$ & $7.2 \times 10^{-14}$ \\
$\super{\sub{\Phi}{ss}}{res}$ & $2.8 \times 10^{-10}$ & $1.4 \times 10^{-13}$
\\ \hline \hline \sub{\Phi}{M31}${}^\mathrm{a}$ &
\multicolumn{2}{c}{$9 \times 10^{-9}$}\\
\end{tabular} 
\end{center}
${}^\mathrm{a}$\citet{2010A&A...523L...2A}
\end{table}%

\section{Discussion}
\label{sec:discussion} 
Thanks to their low velocity dispersion and relatively high density, cold tidal streams and young caustics can provide a significant boost to the dark matter self-annihilation rate if the cross section is non-linearly dependent on the relative velocity, as in the Sommerfeld scenario.  The particular morphology of tidal streams, their location far from the centres of galaxies, and the apparent tracking of the stellar and dark components also make these features an attractive place to search for an annihilation signal, as the correlation with the stellar shape makes it easier to differentiate such a signal from a smooth halo distribution.  In fact, at distances typical of tidal debris 30-50\% of the halo's mass may be in streams \citep{2010arXiv1010.2491M,2011MNRAS.413.1373W}.

In this work we computed the boost to a smooth background from tidal substructure observed in star-count maps of the Andromeda galaxy for two different velocity-dependent cross sections proposed for Sommerfeld-type annihilations between weakly interacting, massive dark matter particles (in which a light helper particle forms in an intermediate state between the dark matter annihilation and the production of standard-model particles).  We used an existing N-body model of the structure to estimate the density and relative velocity of the material in the substructure with suitably unbiased algorithms and a reasonable choice of numerical parameters.  We found that in both cases, the emission from the tidal structure could locally boost the annihilation emission by up to factors of 5.  The case where the cross section $\langle \sigma v\rangle \propto 1/v^2$ produced a boost large enough for the signal from the tidal structure to outshine the smooth halo at large radii, though the estimated signal is several orders of magnitude below the current {\it Fermi} sensitivity for reasonable choices of the dark matter parameters.  However, if an instrument with the required sensitivity existed, a search for emission from the tidal substructure would constitute a test of the velocity-dependence of the dark matter cross section, since only in the $1/v^2$ case is the tidal structure visible.

\section{Future work}
\label{sec:future}

One intriguing result of this work is that tidal streams that are sufficiently massive and collimated (i.e., young and/or cold) can produce significant local boosts of the annihilation signal if the cross-section is velocity-dependent.  Such streams could be a significant contributor to the gamma-ray luminosity of the outer regions of haloes in the Sommerfeld scenario, especially if cosmological simulations accurately predict the percentage of streams.  State-of-the-art cosmological simulations of individual galactic haloes can resolve this coarse-grained phase space structure, and could be used to estimate this contribution.

Low-frequency radio observations could also be used to search for signals from dark matter in tidal substructures, via channels that produce high-energy electrons that then give rise to synchrotron radiation through interactions with the galactic magnetic field.  A map of the polarisation must be correlated with the observed stellar stream, under the assumption that the dark matter and stars track each other, to search for such a signal (Zaroubi, private communication). In future work, we will consider whether this channel could produce a signal detectable with a low-frequency array such as LOFAR, either in M31 or in high-latitude streams in our own Galaxy.

\section{Acknowledgements}
The work for this paper was done during a research stay at the Institut d'Astrophysique de Paris by R.E.S., who thanks the Institut for their gracious hospitality. R.M. thanks French ANR OTARIE for support.

\bibliography{Sommerfeld}

\end{document}